\documentclass[a4paper,prd,preprint,amsmath,
               superscriptaddress,nofootinbib,floatfix,showpacs]{revtex4}

\usepackage{graphicx}

\begin{document}

\title{Exploring flavor structure of supersymmetry breaking from rare
$B$ decays and unitarity triangle}
\date{12/8/2003}

\author{Toru Goto}
\altaffiliation[Present address: ]{YITP, Kyoto University, Kyoto 606-8502, Japan}
\email{goto@yukawa.kyoto-u.ac.jp}
\affiliation{Department of Physics, Graduate School of Science, 
             Osaka University, Toyonaka, Osaka 560-0043, Japan}

\author{Yasuhiro Okada}
\email{yasuhiro.okada@kek.jp}
\affiliation{Theory Group, KEK, Tsukuba, Ibaraki 305-0801, Japan}
\affiliation{Department of Particle and Nuclear Physics,
             The Graduate University of Advanced Studies,
             Tsukuba, Ibaraki 305-0801, Japan}

\author{Yasuhiro~Shimizu}
\email{shimizu@eken.phys.nagoya-u.ac.jp}
\affiliation{Department of Physics, Nagoya University, Nagoya 464-8602, Japan}

\author{Tetsuo Shindou}
\email{shindou@het.phys.sci.osaka-u.ac.jp}
\author{Minoru Tanaka}
\email{tanaka@phys.sci.osaka-u.ac.jp}
\affiliation{Department of Physics, Graduate School of Science, 
             Osaka University, Toyonaka, Osaka 560-0043, Japan}

\begin{abstract}
We study effects of supersymmetric particles in various rare $B$ decay 
processes as well as in the unitarity triangle analysis. We consider
three different supersymmetric models,
the minimal supergravity, SU(5) SUSY GUT with right-handed neutrinos,
and the minimal supersymmetric standard model with U(2) flavor symmetry.
In the SU(5) SUSY GUT with right-handed neutrinos,
we consider two cases of the mass matrix of the right-handed neutrinos.
We calculate direct and mixing-induced CP asymmetries in the 
$b\to s\gamma$ decay and CP asymmetry in $B_d\to \phi K_S$ as
well as the $B_s$--$\bar{B}_s$ mixing amplitude for the unitarity 
triangle analysis in these models.
We show that large deviations are possible for the SU(5) SUSY GUT and the
U(2) model. 
The patterns and correlations of deviations
from the standard model will be useful to discriminate
the different SUSY models in future $B$ experiments.
\end{abstract}


\preprint{OU-HET 443}
\preprint{KEK-TH-885}
\preprint{DPNU-03-14}
\preprint{hep-ph/0306093}

\maketitle
\section{Introduction}
Success of $B$ factory experiments at KEK and SLAC indicates that
$B$ physics is important to explore origins of 
the flavor mixing and the CP violation in the quark sector.
The CP asymmetry in the $B_d\to J/\psi K_S$ mode is precisely
determined, and  a CP violation parameter, 
$\sin 2\phi_1$ (or $\sin 2\beta$) is found to be 
consistent with the prediction in the
standard model (SM) \cite{belleandbabar}. 
In addition, branching ratios and CP asymmetries
in various rare $B$ decays have been reported.
In future, we expect much improvement in measurements of CP
violation and rare $B$ decays at $e^+e^-$ $B$ factories
as well as hadron $B$ experiments \cite{hadronb}.

Goals of future $B$ physics are not only to very precisely determine 
the parameters of the
Cabbibo-Kobayashi-Maskawa (CKM) matrix elements \cite{km}, but also
to search for new sources of CP violation and flavor mixings.
For instance, scalar quark (squark) mass matrices could be such new sources 
in supersymmetric 
models. Since the flavor structure of the squark mass matrices depends on 
the mechanism of supersymmetry (SUSY) breaking at a higher energy scale and
interactions above the weak scale through the renormalization,
future $B$ physics can provide quite important information on the
origin of SUSY breaking.

In our previous papers \cite{gosst1,gosst_proc}, 
we studied the flavor signals in three different
models namely, the minimal supergravity (mSUGRA), 
the SU(5) SUSY GUT with right-handed
neutrinos, and the minimal supersymmetric standard model (MSSM) with U(2)
flavor symmetry \cite{u2-1,u2-2}. 
These models are typical solutions of the SUSY flavor problem.
They have different flavor structures in the squark mass matrices at
the electroweak scale.
Thus, they may be distinguished by low energy quark flavor signals.
We calculated SUSY contributions to the $B_d$--$\bar{B}_d$,
$B_s$--$\bar{B}_s$, and $K^0$--$\bar{K}^0$ mixings in these
models, and showed that the consistency test of the unitarity triangle
from angle and length measurements are useful to discriminate 
these models.

In addition to the consistency test of the unitarity triangle, there are several
promising ways to search for new physics effects through $B$ decay 
processes. As pointed out in the context of SUSY models
\cite{th_bphiks3,th_bphiks1}, 
comparison of time-dependent CP asymmetries in $B_d\to J/\psi K_S$,
$B_d\to \phi K_S$, and $B_d\to \eta'K_S$ provides us with information on new CP
phases in the $b\to s$ transition, because these asymmetries are expected
to be the same in the SM.
Recent results on the CP asymmetry in $B_d\to \phi K_S$ by Belle and 
BaBar collaborations indicate a 2.7$\sigma$
deviation from the SM prediction \cite{exp_bphiks}.
This anomaly may be attributed
to new physics, SUSY with $R$ parity \cite{bphiks_susy_nu,bphiks_susy_rp},
SUSY without $R$ parity \cite{bphiks_susy_wo_rp},
or other models \cite{bphiks_np}.
Moreover, the CP asymmetries in the $b\to s\gamma$ process 
\cite{directcpbsgamma,mixcpbsgamma} have been extensively studied
in several models, and they would exhibit substantial deviation
from the SM in some models \cite{cpv_in_bsgamma}.
The branching
ratio and decay distributions of $b\to sl\bar{l}$ 
have also examined by several authors in different contexts of new physics, 
and they may probe 
different aspects of new physics 
\cite{past_works_bsll_0,past_works_bsll}.

In this paper, we extend our previous analysis \cite{gosst1} 
to rare $B$ decay processes, 
especially
$b\to s$ transitions.
We investigate the direct CP asymmetry in $b\to s\gamma$, and the mixing
induced CP violation in $B_d\to M_s\gamma$ process, where $M_s$ is a
CP eigenstate hadron with strangeness,
and CP asymmetry in $B_d\to \phi K_S$ process.
We show that the above three models exhibit 
different patterns of deviation from the SM.
These observables as well as those
studied in our previous paper \cite{gosst1}
could play an important role in new physics search
at future $B$ experiments such as a super $B$ factory \cite{superB}
and hadron $B$ experiments \cite{hadronb}.


The strategy of the present work and \cite{gosst1} is different from 
most of other works.
For each model of SUSY, we calculate SUSY effects in various
mixings ($B_d$--$\bar{B}_d$, $B_s$--$\bar{B}_s$, and $K^0$--$\bar{K}^0$)
and rare decay processes, and identify  possible patterns of  deviations from
the SM predictions. We then compare patterns of the new 
physics signals for different models. In this way we may be able to 
distinguish different models of SUSY breaking scenarios, or at least
obtain important clues to identity the SUSY breaking sector. Most of 
past works deal with a specific observable signal in a particular SUSY
model. The strategy of combining various information in $B$ physics 
will be important in future, especially in the days of a super $B$ factory 
and dedicated hadron $B$ experiments. The purpose of the present work is
to demonstrate how such global analysis in $B$ physics is useful to explore
the flavor structure of SUSY breaking sectors.

In the present work, some of our calculations are re-analysis of
past works under most up-dated phenomenological constraints, and
others are new studies. Even in the case that a similar calculation
can be found in the literature for a specific process and a model, 
we repeat the calculation in order to treat various processes 
in a uniform way. In the mSUGRA model,   
$B_d$--$\bar{B}_d$, $B_s$--$\bar{B}_s$, and $K^0$--$\bar{K}^0$ 
mixing, and the direct CP violation of $b\to s\gamma$ were 
studied previously. Although the mixing-induced CP violation in 
$B_d\to M_s\gamma$ process and CP asymmetry in $B_d\to \phi K_S$ process
have not explicitly considered before, it was recognized that such processes
did not induce interesting signals once the severe constraints from 
various electric dipole moments were applied. We have confirmed 
this by explicit calculations. For the SU(5) SUSY GUT with right-handed
neutrinos, we have found that flavor signals are quite different 
for different choices of the heavy right-handed neutrino mass matrix.
The ``degenerate case'' defined in the next section was considered
in \cite{susy_gut_with_nu1, susy_gut_with_nu2, gosst1}.
In particular, SUSY effects on the mixing amplitudes were studied
in \cite{susy_gut_with_nu1, susy_gut_with_nu2}, and the 
mixing-induced CP violation in $B_d\to M_s\gamma$ process
was also considered in \cite{susy_gut_with_nu1}. The direct CP violation
in $b\to s\gamma$ and the CP asymmetry in $B_d\to \phi K_S$ process
are analyzed here for the first time. The analysis of $B$ physics signals in 
``non-degenerate case'' is new. We compare the implications of consistency
test of the unitarity triangle for degenerate and non-degenerate cases,
which has been done only for degenerate case in \cite{gosst1}.
We also present analysis of rare decay processes and the $B_s$
mixing for non-degenerate case, which partly overlaps with a recent work
in \cite{bphiks_susy_nu}. For the U(2) model, flavor signals has been 
considered previously for mixing amplitudes in \cite{gosst1,Masiero:2001cc}, 
but a quantitative analysis of various 
rare decay modes has not yet appeared in the literature.
Preliminary results of our analysis are presented in a workshop report
\cite{gosst_proc}, but the non-degenerate case is not included 
in Ref.~\cite{gosst_proc}.

Although we tried to clarify a characteristic feature of each model, 
this cannot be complete because of large numbers of free parameters.
There might be some parameter space where new contributions
become particularly important, which are not shared in a generic
parameter space. For example, it was pointed out recently that neutral 
Higgs boson exchange effects may contribute to the various flavor changing 
process, especially for a large value of the ratio of the two vacuum 
expectation values ($\tan{\beta}$) \cite{neut_higgs_bb}, but we do not take 
into account these effects. Although we estimate that such an effect is small
for the parameter region presented numerically in this paper, this 
effect will be potentially important. 

This paper is organized as follows. In Sec.~II, we introduce the three 
models. The $B_d$--$\bar{B}_d$ mixing, 
the $B_s$--$\bar{B}_s$ mixing, CP asymmetries
in $b\to s\gamma$ and $B_d\to \phi K_S$ are discussed in Sec.~III.
The numerical results on these observables are presented in Sec.~IV.
Our conclusion is given in Sec.~V.

\section{Models}
In this section, we give a brief review of the models.
They are well-motivated examples of SUSY models,
and are chosen as representatives
that have distinct flavor structures.
A detailed description of these models can be found in Ref.~\cite{gosst1}.

\subsection{The minimal supergravity model}
In the mSUGRA, SUSY is spontaneously broken in the hidden 
sector and our MSSM sector is only connected to the
hidden sector by the gravitation.
The soft breaking terms
are induced through the gravitational interaction, and the
soft breaking terms have no new flavor mixing
at the scale where they are induced.

The soft breaking terms are specified at the GUT scale by 
the universal scalar mass ($m_0$), the universal gaugino mass ($M_{1/2}$), 
and the universal trilinear coupling ($A_0$). 
The soft breaking terms at the electroweak scale are determined
by solving renormalization group equations.

In this model, the only source of flavor mixings is the CKM matrix.
New flavor mixings at the electroweak scale come from 
the CKM matrix through radiative corrections. 

As for CP violation, in addition to the CP phase in the CKM matrix,
we have two new CP phases.
One is the complex phase of the $\mu$ term ($\phi_{\mu}$),
and another is the phase of $A_0$ ($\phi_A$).
Since the potential sensitivity of the neutron electric dipole moment (EDM)
to CP violations in low energy SUSY models was stressed
by Ellis {\em et al.} in Ref.~\cite{edm1},
the neutron EDM and the electron EDM have been studied in detail by
several authors
in different context of low energy SUSY \cite{edm1,edm2,edm3}.
The bottom line in the mSUGRA is
that $\phi_{\mu}$ and $\phi_A$ contribute to the neutron and 
the electron EDM's,
and experimental constraints \cite{exp_edmn,exp_edme}  
on these phases are very severe.
Taking these EDM constraints into account, effects of new CP phases
on $K$ and $B$ physics have turned out to be small \cite{th_phase_kb}.

\subsection{The SU(5) SUSY GUT with right-handed neutrinos}
In the last decade, three gauge coupling constants were determined
precisely at LEP and other experiments, and the measured values
turned out to be consistent with the prediction of supersymmetric 
grand unification.
Furthermore, recent developments of neutrino experiments established
the existence of small finite masses of neutrinos 
\cite{exp_sol_atm_nu,exp_k2k,exp_kamland}, which can be
naturally accommodated by the seesaw model \cite{seesaw}.
Guided by these observations, we consider SU(5) SUSY GUT 
with right-handed neutrinos.

In this model, the soft breaking terms are the same as in the mSUGRA model
at the scale where the soft breaking terms are induced. Unlike the 
mSUGRA model, the SU(5) SUSY GUT with right-handed neutrinos
has new sources of flavor mixing in the neutrino sector.
A large flavor mixing in the neutrino 
sector can affect the right-handed down type squark sector
through GUT interactions. Quark flavor signals in this model
have been studied in 
Ref.~\cite{bphiks_susy_nu,th_bphiks1,susy_gut_with_nu1,susy_gut_with_nu2,susy_so10gut_with_nu}.

In the seesaw model, the neutrino mass matrix is written as
\begin{align}
(m_{\nu})^{ij}=\langle h_2\rangle^2(y_{\nu})^{ki}
(M_N^{-1})^{kl}(y_{\nu})^{lj}\;,
\end{align}
where $y_{\nu}$ is the neutrino Yukawa coupling constant matrix, 
$M_N$ is the mass matrix of right-handed neutrinos, 
$\langle h_2\rangle$ denotes the vacuum expectation value 
of one of the Higgs fields $h_2$, and
$i,j,k,l$ are generation indices.
In the basis that the charged lepton Yukawa coupling constant matrix 
is diagonal, this neutrino mass
matrix is related to the observable neutrino mass eigenvalues and the 
Maki-Nakagawa-Sakata (MNS) matrix \cite{mns} as
\begin{align}
(m_{\nu})^{ij}=
(V_{\text{MNS}}^*)^{ik}m_{\nu}^k(V_{\text{MNS}}^{\dagger})^{kj}\;.
\end{align}

In this model, the scalar lepton (slepton) masses and the $A$ terms have
the mSUGRA type structure at the Planck scale ($M_\text{P}$) as
\begin{align}
(m_L^2)^{ij}=m_0^2\delta^{ij}\;,\quad
(m_E^2)^{ij}=m_0^2\delta^{ij}\;,\quad
(A_E)^{ij}=m_0A_0y_e^{i}\delta^{ij}\;,
\end{align}
where $m_L^2$ and $m_E^2$ are the mass squared matrices of sleptons and 
$A_E$ denotes the slepton trilinear scalar couplings.
However, at the scale $M_R$, where the right-handed Majorana neutrinos are
decoupled, new flavor mixings are generated by the renormalization
group effects \cite{neutrino_muegamma0}.
In the leading logarithmic approximation, they are given as
\begin{subequations}
\begin{align}
(m_L^2)^{ij}\simeq& -\frac{1}{8\pi^2}m_0^2(3+|A_0|^2)
(y_{\nu}^{\dagger}y_{\nu})^{ij}\ln \frac{M_{\text{P}}}{M_R}\;,\\
(m_E^2)^{ij}\simeq& 0\;,\\
(A_E)^{ij}\simeq&
-\frac{3}{8\pi^2}m_0A_0y_e^i(y_{\nu}^{\dagger}y_{\nu})^{ij}
\ln \frac{M_{\text{P}}}{M_R}\;,
\end{align}
\end{subequations}
for $i\neq j$. Consequences of these mixings on lepton flavor
violating processes have been investigated from various aspects
\cite{neutrino_muegamma1,neutrino_muegamma2}. We see that,
in this model, lepton flavor violating processes 
such as $\mu\to e\gamma$ are sensitive to the off-diagonal 
elements of $y_{\nu}^{\dagger}y_{\nu}$.

We consider two cases in the SU(5) SUSY GUT with right-handed 
neutrinos in regard to the spectrum of the 
right-handed Majorana neutrinos. One is the case that
all the masses of heavy Majorana neutrinos are the same (degenerate case).
In this simplest case, the flavor mixing of the neutrino sector is only caused
by $y_{\nu}$ because there is no flavor 
mixing in $M_N$. Since the large mixing in the MNS matrix implies that
the off-diagonal elements of $y_{\nu}$ is large,
the $\mu\to e\gamma$ branching ratio is enhanced in the wide region 
of the parameter space and exceeds the experimental bound in some
parameter regions.
In order to suppress the $\mu\to e\gamma$ branching ratio,
we consider an elaborated case that
$M_N$ is not proportional to the unit
matrix (non-degenerate case). In this case, 
the neutrino mixing comes from both $y_{\nu}$ and $M_N$.
If the large mixing in the MNS matrix originates from $M_N$,
the corresponding off-diagonal element of $y_{\nu}$ need not
be large.
Thus, the $\mu \to e\gamma$ decay rate can be suppressed in this case.

We have new CP phases in this model. They are classified
in the following three classes:

{
\def\labelenumi{(\roman{enumi})}
\begin{enumerate}
\item The CP phases in the mSUGRA, \emph{i.e.}, $\phi_{\mu}$ and $\phi_A$.
\item CP phases in the neutrino sector.
There are six physical complex phases 
in $y_{\nu}$ and $M_N$
in the basis in which the charged lepton mass matrix is real and
diagonal. From the combination of these six CP phases, we obtain
three CP phases
in the low energy region, \emph{i.e.},
one Dirac CP phase and two Majorana CP phases.
\item GUT CP phases \cite{gutcp,susy_gut_with_nu2,th_bphiks1}.
The quark and lepton superfields are embedded in $\pmb{10}$ and
$\bar{\pmb{5}}$ representations of SU(5) as
\begin{align}
\pmb{10}_i=
\left\{Q_i, e^{-i\phi_i^Q}(V^{\dagger}\bar{U})_i,
e^{i\phi_i^L}\bar{E}_i\right\}\;,
\quad
\bar{\pmb{5}}_i=
\left\{\bar{D}_i, e^{-i\phi_i^L}L_i\right\}\;,
\end{align}
where $V$ is the CKM matrix at the GUT scale,
$Q_i(\pmb{3},\pmb{2},1/6)$, 
$\bar{U}_i(\bar{\pmb{3}},\pmb{1},-2/3)$,
$\bar{D}_i(\bar{\pmb{3}},\pmb{1},1/3)$,
$L_i(\pmb{1},\pmb{2},-1/2)$, and
$\bar{E}_i(\pmb{1},\pmb{1},1)$
are quark and lepton superfields in the $i$th generation with
the $\text{SU(3)}\times \text{SU(2)}\times \text{U(1)}$ gauge quantum numbers
in parentheses. The phases $\phi_i^L$ and $\phi_i^Q$ obey the 
constraints $\phi_1^L+\phi_2^L+\phi_3^L=0$ and 
$\phi_1^Q+\phi_2^Q+\phi_3^Q=0$.
Before the SU(5) is broken, CP phases $\phi^L_i$ and $\phi_i^Q$ have
physical meanings and they may play an important role
in the flavor physics
through the renormalization group effect above the GUT scale.
\end{enumerate}
}

\subsection{A model with U(2) flavor symmetry}
An alternative solution of the flavor problem of SUSY is
introducing some flavor symmetry. U(2) flavor symmetry is
one of such symmetries \cite{u2-1,u2-2}. 
We consider the model given in Ref.~\cite{u2-2}.
In this model, the quark and lepton supermultiplets in the first and 
the second generations transform as doublets under the U(2) flavor 
symmetry, and the third generation and the Higgs supermultiplets are
singlets under the U(2).
 
In order to reproduce the correct structure of the quark Yukawa 
coupling matrices, we 
assume the following breaking pattern of the U(2):
\begin{align}
\mbox{U(2)}\rightarrow \mbox{U(1)}\rightarrow \pmb{1}(\text{no symmetry})\;.
\end{align}
With this assumption, we obtain the quark Yukawa coupling matrix $y_Q$
and the squark mass matrices $m_X^2$:
\begin{align}
y_Q^{ij}=Y_Q
\begin{pmatrix}
0&a_Q\epsilon^{\prime}&0\\
-a_Q\epsilon^{\prime}&d_Q\epsilon&b_Q\epsilon\\
0&c_Q\epsilon&1
\end{pmatrix}\;,
\quad Q=U,D\;,
\label{eq_u2_yukawa}
\end{align}
\begin{align}
m_X^2=(m_0^{X})^2
\begin{pmatrix}
1&0&0\\
0&1+r_{22}^{X}\epsilon^2&r_{23}^X\epsilon\\
0&r_{23}^{X*}\epsilon&r_{33}^X
\end{pmatrix}\;,
X=Q,U,D\;,
\label{eq_u2_susy_para}
\end{align}
where $\epsilon$ and 
$\epsilon^{\prime}$ are order parameters of the U(2) and U(1)
symmetry breaking respectively and they satisfy 
$\epsilon^{\prime}\ll \epsilon\ll 1$, and 
$Y_Q$, $a_Q$, $b_Q$, $c_Q$, $d_Q$, and $r^X$ are
dimensionless parameters of $\mathcal{O}(1)$.
As for the squark $A$ terms, they have the same structure as
the quark Yukawa coupling matrices:
\begin{align}
A_Q^{ij}=A_Q^0Y_Q
\begin{pmatrix}
0&\tilde{a}_{Q}\epsilon^{\prime}&0\\
-\tilde{a}_{Q}\epsilon^{\prime}&\tilde{d}_Q\epsilon&
\tilde{b}_Q\epsilon\\
0&\tilde{c}_Q\epsilon&1
\end{pmatrix}\;,\quad
Q=U,D\;.
\end{align}
In general, though being of $\mathcal{O}(1)$, 
$\tilde{a}_Q$, $\tilde{b}_Q$, $\tilde{c}_Q$, and
$\tilde{d}_Q$ take different values from the
corresponding parameters in Eq.~(\ref{eq_u2_yukawa}), and
we expect no exact universality of the $A$ terms in this model.

In the mass matrices of sfermions in this model,
the degeneracy between masses of the first and the second generation
is naturally realized. On the other hand, the mass of the third generation may
be separated from the others. There exist flavor mixings of 
$\mathcal{O}(\epsilon)$ between
the second and the third generations of sfermions. These are new sources
of flavor mixing besides the CKM matrix.

There are several efforts to explain the observed neutrino masses and
mixings in SUSY models with the U(2) flavor symmetry
(or its discrete relatives)
\cite{neutrino_u2}. 
However, the purpose of this paper is to illustrate
typical quark, especially bottom, flavor signals in several
SUSY models and to examine the possibility to distinguish them.
A detailed study of neutrino masses and mixings and lepton flavor signals
in models with flavor symmetry is beyond the scope of the present work.
Hence, in the following analysis, we will not consider the lepton sector
in the U(2) model.

\section{Processes}
\label{sec_processes}
The processes considered in the following are 
the $B_d$--$\bar{B}_d$ and the $B_s$--$\bar{B}_s$ mixings, $b\to s\gamma$,
and $B\to \phi K_S$.
We consider the effective Lagrangian that consists of
$\Delta B=1$ terms and $\Delta B=2$ terms:
\begin{align}
\mathcal{L}=\mathcal{L}_{\Delta B=1}+\mathcal{L}_{\Delta B=2}\;,
\end{align}
\begin{align}
\mathcal{L}_{\Delta B=1}=&
C_{2L}\mathcal{O}_{2L}
+C_{2L}^{\prime}\mathcal{O}_{2L}^{\prime}
+C_{LL}\mathcal{O}_{LL}
+C_{LR}^{(1)}\mathcal{O}_{LR}^{(1)}
+C_{LR}^{(2)}\mathcal{O}_{LR}^{(2)}
+C_{TL}^{(1)}\mathcal{O}_{TL}^{(1)}
+C_{TL}^{(2)}\mathcal{O}_{TL}^{(2)}\nonumber\\
&-C_{7L}\mathcal{O}_{7L}
-C_{8L}\mathcal{O}_{8L}
+(L\leftrightarrow R)
\end{align}
where 
$\mathcal{O}$'s are
\begin{subequations}
\begin{align}
\mathcal{O}_{2L}=&
(\bar{s}_{\alpha}\gamma^{\mu}c_{L\alpha})
(\bar{c}_{\beta}\gamma^{\mu}b_{L\beta})\;,\\
\mathcal{O}_{2L}^{\prime}=&
(\bar{s}_{\alpha}\gamma^{\mu}u_{L\alpha})
(\bar{u}_{\beta}\gamma^{\mu}b_{L\beta})
-(\bar{s}_{\alpha}\gamma^{\mu}c_{L\alpha})
(\bar{c}_{\beta}\gamma^{\mu}b_{L\beta})
\;,\\
\mathcal{O}_{LL}=&
(\bar{s}_{\alpha}\gamma^{\mu}b_{L\alpha})
(\bar{s}_{\beta}\gamma_{\mu}s_{L\beta})\;,\\
\mathcal{O}_{LR}^{(1)}=&
(\bar{s}_{\alpha}\gamma^{\mu}b_{L\alpha})
(\bar{s}_{\beta}\gamma_{\mu}s_{R\beta})\;,\\
\mathcal{O}_{LR}^{(2)}=&
(\bar{s}_{\alpha}\gamma^{\mu}b_{L\beta})
(\bar{s}_{\beta}\gamma_{\mu}s_{R\alpha})\;,\\
\mathcal{O}_{TL}^{(1)}=&
\frac{1}{4}(\bar{s}_{\alpha}[\gamma^{\mu},\gamma^{\nu}]b_{L\alpha})
(\bar{s}_{\beta}[\gamma_{\mu},\gamma_{\nu}]s_{L\beta})\;,\\
\mathcal{O}_{TL}^{(2)}=&
\frac{1}{4}(\bar{s}_{\alpha}[\gamma^{\mu},\gamma^{\nu}]b_{L\beta})
(\bar{s}_{\beta}[\gamma_{\mu},\gamma_{\nu}]s_{L\alpha})\;,\\
\mathcal{O}_{7L}=&\frac{e}{16\pi^2}m_b\bar{s}
\frac{i}{2}[\gamma^{\mu},\gamma^{\nu}]b_RF_{\mu\nu}\;,\\
\mathcal{O}_{8L}=&\frac{g_3}{16\pi^2}m_b\bar{s}^{\alpha}
\frac{i}{2}[\gamma^{\mu},\gamma^{\nu}]
T_{\alpha\beta}^{(a)}b_R^{\beta}G_{\mu\nu}^{(a)}\;.
\end{align}
\end{subequations}
Among the above Wilson coefficients, $C_{2L}$ is dominated by the
tree contributions from the SM at the weak scale 
and the others are induced by loop effects. Therefore, one obtains
$C_{2L}^{\prime}=\epsilon_u C_{2L}$, where
$\epsilon_u=-V^*_{us}V_{ub}/V^*_{ts}V_{tb}$.
The coefficients $C_{LL}$, $C_{LR}^{(1)}$, and $C_{LR}^{(2)}$
are also dominated by the SM contribution because of the QCD correction
below the electroweak scale.
$\mathcal{L}_{\Delta B=2}$ is
described in our previous paper \cite{gosst1}.

We discussed 
the $B_d$--$\bar{B}_d$ and
the $B_s$--$\bar{B}_s$ mass splittings $\Delta m_{B_d}$ and $\Delta m_{B_s}$
in Ref.~\cite{gosst1}.
In this paper, 
we consider both the direct
and the mixing induced CP asymmetries in $b\to s\gamma$ and 
the CP asymmetry in
$B\to \phi K_S$ in addition to the above $\Delta B=2$ process.

About the $B^0$--$\bar{B}^0$ mixing processes, 
the mixing matrix elements $M_{12}(B_d)$ and $M_{12}(B_s)$
are defined as
\begin{align}
M_{12}(B_q)=&
-\frac{1}{2m_{B_q}}\langle B_q|\mathcal{L}_{\Delta B=2}|\bar{B}_q\rangle\;,
\end{align}
where $q=d,s$. We can express $\Delta m_{B_d}$
and $\Delta m_{B_s}$ in terms of $M_{12}(B_q)$ as
\begin{align}
\Delta m_{B_q}=&2|M_{12}(B_q)|\;.
\end{align}

The direct CP asymmetry in the inclusive decays $B\to X_s\gamma$ 
is defined as \cite{directcpbsgamma}
\begin{align}
A_{CP}^{\text{dir}}(B\to X_s\gamma)=&
\frac{\Gamma (\bar{B}\to X_s\gamma)-\Gamma(B\to X_{\bar{s}}\gamma)}
{\Gamma (\bar{B}\to X_s\gamma)+\Gamma(B\to X_{\bar{s}}\gamma)}\nonumber\\
=&
-\frac{\alpha_3}{\pi(|C_{7L}|^2+|C_{7R}|^2)}
\Biggl[
-\mathrm{Im}r_2\mathrm{Im}\left[(1-\epsilon_u)C_{2L}C_{7L}^*\right]
+\frac{80}{81}\pi\mathrm{Im}(\epsilon_uC_{2L}C_{7L}^*)\nonumber\\
&+\frac{8}{9}\pi\mathrm{Im}(C_{8L}C_{7L}^*)
-\mathrm{Im}f_{27}
\mathrm{Im}\left[(1-\epsilon_u)C_{2L}C_{7L}^*\right]\nonumber\\
&+\frac{1}{3}\mathrm{Im}f_{27}\mathrm{Im}
\left[(1-\epsilon_u)C_{2L}C_{8L}^*\right]
+(L\leftrightarrow R)
\Biggr]\;,
\label{eq_dircpbsgamma}
\end{align}
where 
the functions $r_2$ and $f_{27}$ are found in Ref.~\cite{cmm-bsgamma}.

The time-dependent CP asymmetry in the $B_d$ decays to a CP eigenstate 
$f_{CP}$ is given by
\begin{align}
\frac{\Gamma(\bar{B}_d(t)\to f_{CP})-\Gamma(B_d(t)\to f_{CP})}
{\Gamma(\bar{B}_d(t)\to f_{CP})+\Gamma(B_d(t)\to f_{CP})}
=&A_{CP}^{\text{dir}}(B_d\to f_{CP})\cos\Delta m_{B_d} t\nonumber\\
&+A_{CP}^{\text{mix}}(B_d\to f_{CP})\sin \Delta m_{B_d} t\;.
\end{align}
In the $b\to s \gamma$ decays,
we consider the time-dependent mixing induced CP asymmetry in
$B_d\to M_s \gamma$, where $M_s$ denotes a hadronic CP eigenstate
which includes a strange quark such as $K^*$ or $K_1$.
$A_{CP}^{\text{mix}}(B_d\to M_s\gamma)$ is given as \cite{mixcpbsgamma}
\begin{align}
A_{CP}^{\text{mix}}(B_d\to M_s\gamma)=
\frac{2\mathrm{Im}(e^{-i\phi_B}C_{7L}C_{7R})}{|C_{7L}|^2+|C_{7R}|^2}\;,
\end{align}
where
\begin{align}
e^{i\phi_B}=\frac{M_{12}(B_d)}{|M_{12}(B_d)|}\;.
\end{align}

As for $B_d\to \phi K_S$, we consider 
$A_{CP}^{\text{mix}}(B_d\to \phi K_S)$, which is given as
\begin{align}
A_{CP}^{\text{mix}}(B_d\to \phi K_S)=
\frac{2\mathrm{Im}(e^{-i\phi_{B}}
\bar{\mathcal{A}}\mathcal{A})}
{|\mathcal{A}|^2+|\bar{\mathcal{A}}|^2}\;,
\end{align}
where $\mathcal{A}$ and $\bar{\mathcal{A}}$ are
decay amplitudes of $B_d\to \phi K$ and $\bar{B}_d\to \phi \bar{K}$ 
respectively.
Since $B_d\to \phi K_S$ is a hadronic decay mode, the calculation
of the decay amplitude suffers from large theoretical uncertainties.
Here we use a method based on the naive factorization.
Details of the calculation of $\mathcal{A}$ is given 
in Refs.~\cite{th_bphiks1,th_bphiks3,th_bphiks4}.
Using the naive factorization ansatz,
we obtain
\begin{align}
\bar{\mathcal{A}}\equiv&
-\langle \phi \bar{K}|\mathcal{L}|\bar{B}\rangle\nonumber\\
=&-H_V\left[\frac{1}{3}C_{LL}+\frac{1}{4}C_{LR}^{(1)}
+\frac{1}{12}C_{LR}^{(2)}-\frac{7}{6}\frac{H_T}{H_V}C_{TL}^{(1)}
-\frac{5}{6}\frac{H_T}{H_V}C_{TL}^{(2)}
+\frac{\alpha_s}{4\pi}\frac{4}{9}\kappa_{\text{DM}}C_{8L}
+(L\leftrightarrow R)\right.\nonumber\\
&\left.
+\frac{1}{9}P_G^{(c)}(q^2,m_b^2)C_{2L}
+\frac{1}{9}\left(P_G^{(u)}(q^2,m_b^2)-P_G^{(c)}(q^2,m_b^2)\right)\epsilon_uC_{2L}
\right]\;,
\end{align}
where $H_V=3\langle \phi \bar{K}|\mathcal{O}_{LL}|\bar{B}\rangle$ and
$H_T=-(6/7)\langle \phi \bar{K}|\mathcal{O}_{TL}^{(1)}|\bar{B}\rangle$.
$P_G^{(u,c)}(q^2,m_b^2)$ comes from the one-loop matrix element
of  $\mathcal{O}_{2L}$,
$q^2$ is the momentum transfer of the exchanged gluon,
and $\kappa_{\text{DM}}$ is an $\mathcal{O}(1)$ coefficient \cite{th_bphiks4}
that parametrizes the matrix element of $\mathcal{O}_8$. 
The concrete form of 
$P_G^{(q_i)}(q^2,m_b^2)$ is 
$P_G^{(q_i)}(q^2,m_b^2)=-(\alpha_s/(4\pi))(G(m_{q_i}^2,q^2,m_b^2)+2/3)$ for the 
NDR scheme \cite{melm_c2} with
\begin{align}
G(m_{q_i}^2,q^2,m_b^2)=-4\int_0^1dxx(1-x)\ln\left(
\frac{m_{q_i}^2-q^2x(1-x)}{m_b^2}\right)\;.
\end{align}
In the constituent quark model, $H_T/H_V$ is proportional to 
$m_{\phi}/m_{B}$,
and thus the contributions from
$C_{TL}^{(1)}$ and $C_{TL}^{(2)}$
are neglected.

In our calculations, we take $q^2=m_b^2/2$ and $\kappa_{\text{DM}}=1$.
Though these approximations are difficult to be justified in QCD,
we employ them for an illustration that may provides
the correct order of magnitude.

\section{Numerical analysis}
\subsection{Parameters}
\subsubsection{Parameters in the minimal supergravity model}
\label{sec_para_msugra}
The parameters we use in our calculation are almost the same as
those used in our previous paper \cite{gosst1} 
except for CP phases. 
In our calculation, we treat the masses and 
the mixing matrices in the quark and lepton sectors
as input parameters that determine the Yukawa coupling matrices.

The CKM matrix elements 
$V_{us}$, 
$V_{cb}$, and $|V_{ub}|$ are determined by experiments
independently of new physics because they are based
on tree level processes.
We adopt $V_{us}=0.2196$ and 
$V_{cb}=0.04$ in the following calculations and
vary $|V_{ub}|$ within a range 
$|V_{ub}/V_{cb}|=0.09\pm 0.01$.
Note that the current error of $|V_{ub}|$ is
estimated to be larger than this value but we expect theoretical and 
experimental improvements in near future. We vary the CKM phase 
$\phi_3\equiv 
\arg (-V^*_{ub}V_{ud}/
V^*_{cb}V_{cd})$ within 
$\pm 180^{\circ}$, because it is not yet constrained by tree level 
processes free from new physics contributions.

As for the SUSY parameters, we take the convention that the unified 
gaugino mass $M_{1/2}$ is real.
It is known that $\phi_{\mu}$ is strongly constrained by the upper
bound of EDM's, while the corresponding constraint on $\phi_A$ 
is not so tight \cite{edm3}.
Accordingly, we fix $\phi_{\mu}$ as 
$0^{\circ}$ or $180^{\circ}$ at the electroweak scale.
We vary the universal scalar mass $m_0$,
$M_{1/2}$, and the proportional constant of $A$ terms to Yukawa coupling 
matrix $A_0m_0$ 
within the ranges
$0<m_0<3\mbox{TeV}$, $0<M_{1/2}<1\mbox{TeV}$, 
$|A_0|<5$, and $-180^{\circ}<\phi_A <180^{\circ}$.
We take the ratio of two VEV's 
$\tan\beta=\langle h_2\rangle/\langle h_1\rangle=30$ or 5.

\subsubsection{Parameters in the SU(5) SUSY GUT with
right-handed neutrinos}

In the SU(5) SUSY GUT with right-handed neutrinos, we need to 
specify the parameters in the neutrino sector in addition to the quark
Yukawa coupling constants given in the previous discussion.
We take the neutrino masses as
$m_{\nu_3}^2-m_{\nu_2}^2=3.5\times 10^{-3}\mbox{eV}^2$,
$m_{\nu_2}^2-m_{\nu_1}^2=6.9\times 10^{-5}\mbox{eV}^2$,
and $m_{\nu_1}\simeq 0.001$eV, and the MNS mixing matrix as
\begin{align}
V_{\text{MNS}}=
\begin{pmatrix}
c_{\odot}c_{13}&s_{\odot}c_{13}&s_{13}\\
-s_{\odot}c_{\text{atm}}-c_{\odot}s_{\text{atm}}s_{13}&
c_{\odot}c_{\text{atm}}-s_{\odot}s_{\text{atm}}s_{13}&
s_{\text{atm}}c_{13}\\
s_{\odot}s_{\text{atm}}-c_{\odot}c_{\text{atm}}s_{13}&
-c_{\odot}s_{\text{atm}}-s_{\odot}c_{\text{atm}}s_{13}&
c_{\text{atm}}c_{13}
\end{pmatrix}\;,
\end{align}
($c_i=\cos\theta_i,s_i=\sin\theta_i$) with $\sin^22\theta_{\text{atm}}=1$,
$\tan^2\theta_{\odot}=0.420$, and $\sin^22\theta_{13}=0$.
These mass squared differences and mixing angles are consistent
with the observed solar and atmospheric neutrino 
oscillations \cite{exp_sol_atm_nu}, the K2K experiment \cite{exp_k2k}, 
and the KamLAND experiment \cite{exp_kamland}.
Only the upper bound of $\sin^22\theta_{13}$ is obtained by
reactor experiments \cite{exp_reac}, 
and we take the above value as an illustration.
We do not introduce the Dirac and Majorana CP phases in the neutrino
sector for simplicity.

We consider $M_R=4.0\times 10^{13}$GeV and
$M_R=4.0\times 10^{14}$GeV for the degenerate case.
(In this case, $M_R$ is the same as the common mass of the right-handed
neutrinos.)
In the non-degenerate case, we take the neutrino Yukawa coupling
matrix as
\begin{align}
y_{\nu}^{\dagger}y_{\nu}
=\begin{pmatrix}
Y_{11}&0&0\\
0&Y_{22}&Y_{23}\\
0&Y_{23}^*&Y_{33}
\end{pmatrix}
\label{y_nu_dagy_nu}
\end{align}
in order to avoid too large SUSY contribution in the 
branching ratio of $\mu\to e\gamma$. The
masses and mixings of the right-handed neutrinos are determined
to reproduce the observed neutrino masses and $V_{\text{MNS}}$.
In Table~\ref{table_nu_para}, we show the numerical values of 
the neutrino parameters in the non-degenerate case.
We integrate out the right-handed neutrinos at $M_R=4.0\times 10^{14}$GeV
in the non-degenerate case.

\begin{table}
\begin{center}
\begin{tabular}{c||c|c}\hline
$\tan\beta$
&$y_{\nu}$&eigenvalues of $M_N$ ($\times 10^{14}$GeV)\\\hline
5&
$\begin{pmatrix}
 0.13&0&0\\
 0&0.099&-0.099\\
 0&0.46&0.46
 \end{pmatrix}$&
\begin{tabular}{c}
$4.4$\\
$0.56$\\
$1.7$\\
\end{tabular}
\\ \hline
30&$\begin{pmatrix}
 0.13&0&0\\
 0&0.098&-0.098\\
 0&0.47&0.47
 \end{pmatrix}$&
\begin{tabular}{c}
$4.5$\\
$0.57$\\
$1.8$\\
\end{tabular}
\\ \hline
\end{tabular}
\end{center}
\caption{The neutrino parameters used in the non-degenerate case.
We show the neutrino Yukawa coupling matrices and
the mass eigenvalues of the right-handed neutrinos
for $\tan\beta=5$ and $\tan\beta=30$. These 
Yukawa coupling matrices give the structure
as given in Eq.~(\ref{y_nu_dagy_nu})
}
\label{table_nu_para}
\end{table}

For the GUT phases $\phi_i^Q$  and $\phi_i^L$, we take
$\phi_i^Q=0$ and vary $\phi_i^L$ within $-180^{\circ}<\phi_i^L<180^{\circ}$
while $\phi_1^L+\phi_2^L+\phi_3^L=0$ is
satisfied.

The soft SUSY breaking parameters in this model are
assumed to be universal at the Planck scale, and
the running effect between the Planck scale and the GUT scale 
is taken into account.
We scan the same ranges for $m_0$, $M_{1/2}$, 
$\phi_A$, and $|A_0|$ as those in the mSUGRA case.

\begin{figure}
\begin{center}
\includegraphics{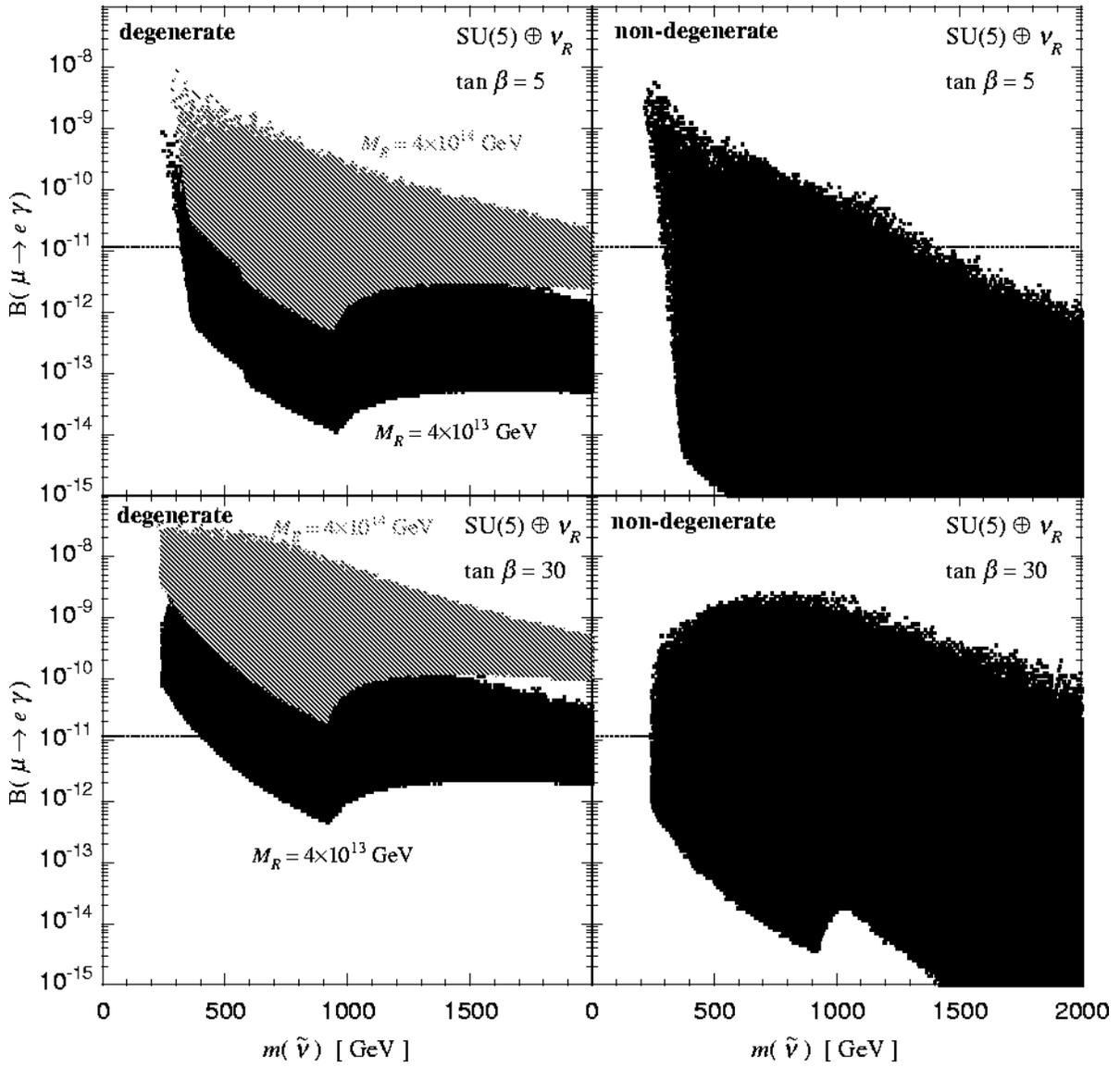}
\end{center}
\caption{
  The branching ratio of $\mu\to e\gamma$ as functions of the
  lightest sneutrino mass in the SU(5) SUSY GUT with right-handed neutrinos.
  The dotted line shows the experimental upper bound.
}
\label{fig_muegamma_su5}
\end{figure}

In Fig.~\ref{fig_muegamma_su5}, we show the branching ratio
of $\mu\to e\gamma$ in both the degenerate and the non-degenerate cases
of the SU(5) SUSY GUT with right-handed neutrinos.
As seen in this figure, the experimental constraint on the parameter
space is more strict for the degenerate case than the
non-degenerate case. 
Both in the degenerate and the non-degenerate cases,
the SUSY contribution to $\mu\to e\gamma$ becomes larger 
for larger $M_R$ \cite{neutrino_muegamma1,neutrino_muegamma2}.
However, the contribution is less significant in the non-degenerate
case for a similar $M_R$ \cite{neutrino_muegamma2}.
In fact, for $M_R=4.0\times 10^{14}\mbox{GeV}$ in
the degenerate case, the SUSY contribution to $\mu\to e\gamma$ is
so large that most of the parameter region is excluded when $\tan\beta=30$.
While, in the non-degenerate case, a large part of the parameter region is
allowed. In the following, we take $M_R=4.0\times 10^{13}\mbox{GeV}$ for
the degenerate case, and 
$M_R=4.0\times 10^{14}\mbox{GeV}$ for the non-degenerate case.

\subsubsection{Parameters in the U(2) model}
In the U(2) model, the symmetry breaking parameters $\epsilon$ and
$\epsilon^{\prime}$ are taken to be $\epsilon=0.04$ and 
$\epsilon^{\prime}=0.008$, and the other parameters in the
quark Yukawa coupling matrices are
determined so that the CKM matrix and the quark masses given in 
Sec.~\ref{sec_para_msugra} are reproduced. The detailed discussion to determine
the quark Yukawa coupling matrices is given in Ref.~\cite{gosst1}.

There are many free parameters in the SUSY breaking sector as shown in 
Eq.~(\ref{eq_u2_susy_para}). In order to reduce the number of free
parameters in numerical calculations, we assume that
\begin{align}
m_0^{Q2}=&m_0^{U2}=m_0^{D2}\equiv m_0^2\;,\\
r_{ij}^{Q}=&r_{ij}^U=r_{ij}^D\equiv r_{ij}\;,\quad (ij)=(22),(23),(33).
\end{align}
We scan the ranges for these parameters as
$0<m_0<3\mbox{TeV}$, $-1<r_{22}<+1$, $0<r_{33}<4$,
$|r_{23}|<4$, and $-180^{\circ}<\arg r_{23}<180^{\circ}$. 
We assume that the boundary conditions for the $A$ parameters
are the same as the mSUGRA case for simplicity. 

\subsection{Experimental constraints}
In order to obtain allowed regions of the parameter space, we consider 
the following experimental results:
\begin{itemize}
\item Lower limits on the masses of SUSY particles and the Higgs
bosons given by direct searches in collider 
experiments \cite{exp_directsearch}.
\item Branching ratio of the $b\to s\gamma$ decay: 
$2\times 10^{-4}<\text{B}(b\to s\gamma)<4.5\times 10^{-4}$ \cite{exp_bsg}.
\item Upper bound of the branching ratio of the $\mu\to e\gamma$ decay 
for the SUSY GUT cases:
$\text{B}(\mu\to e\gamma)<1.2\times 10^{-11}$ \cite{exp_mueg}.
\item Upper bounds of EDM's of the neutron and the electron: 
$|d_n|<6.3\times 10^{-26}e\cdot\text{cm}$ \cite{exp_edmn}
and 
$|d_e|<4.0\times 10^{-27}e\cdot\text{cm}$ \cite{exp_edme}.
\item The CP violation parameter $\varepsilon_K$ in the $K^0$--$\bar{K}^0$
mixing
and the $B_d$--$\bar{B}_d$ mixing parameter $\Delta m_{B_d}$ \cite{exp_mbd}.
As for the $B_s$--$\bar{B}_s$ mixing parameter $\Delta m_{B_s}$, we take 
$\Delta m_{B_s}>13.1\mbox{ps}^{-1}$ \cite{exp_mbs}.
\item CP asymmetry in the $B\to J/\psi K_S$ decay and related modes 
observed at the B factory experiments \cite{belleandbabar}.
\end{itemize}

\subsection{Numerical results}
\subsubsection{Unitarity triangle analysis}
\begin{figure}
\begin{center}
\includegraphics{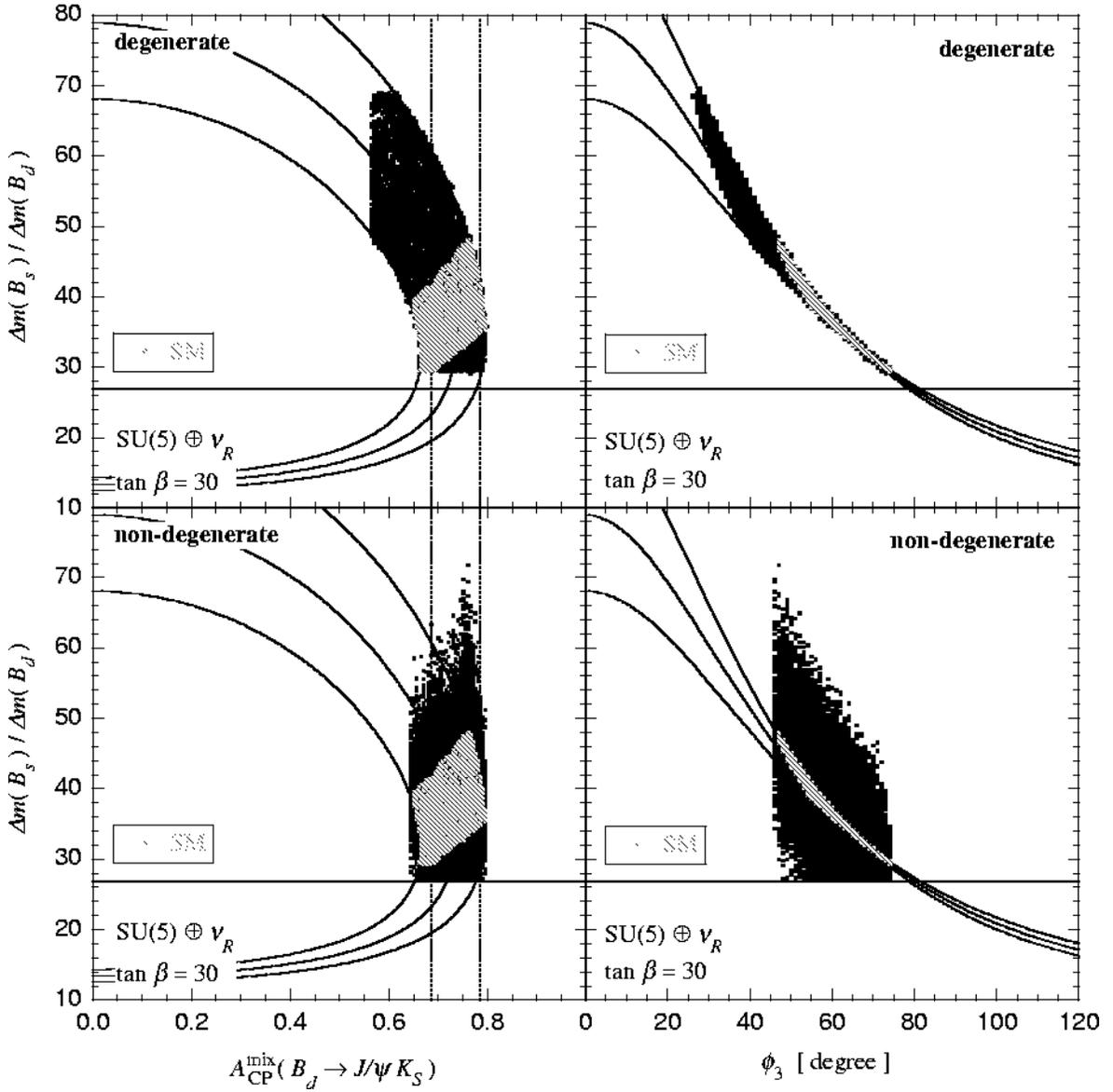}
\end{center}
\caption{
  $\Delta m_{B_s}/\Delta m_{B_d}$ versus the mixing-induced CP asymmetry
  of $B_d\to J/\psi K_S$ and $\phi_3$ in the SU(5) SUSY GUT with
  right-handed neutrinos.
  The light-colored regions show the allowed region in the SM.
  The curves show the SM values with $|V_{ub}/V_{cb}|=0.08$, 0.09
  and 0.10.
  This plot corresponds to Fig.~5 of Ref.~\cite{gosst1}.
}
\label{fig_dmbsd_acppsiks}
\end{figure}

As in our previous work \cite{gosst1}, we search possible values of 
$A_{CP}^{\text{mix}}(B_d\to J/\psi K_S)$, $\Delta m_{B_s}$, $\Delta m_{B_d}$,
and $\phi_3$ under the constraints stated above.
The results for the mSUGRA and the U(2) model are not 
shown here, since they
are similar as Fig.~5 in Ref.~\cite{gosst1} apart from 
slight changes in some input parameters.

In Fig.~\ref{fig_dmbsd_acppsiks}, 
we show the above quantities in the non-degenerate
case of the SU(5) SUSY GUT with right-handed neutrinos for $\tan\beta=30$
as well as the degenerate case.
In the degenerate case, as we found in our previous paper \cite{gosst1},
the SUSY contribution to the $K^0$--$\bar{K}^0$ mixing is large
and $\varepsilon_K$ is
significantly affected. This is because of the large 1--2 mixing in the squark
sector. However, the SUSY contributions to the $B_d$--$\bar{B}_d$
mixing and the $B_s$--$\bar{B}_s$ mixing are not important.
Thus the correlation
among $\Delta m_{B_s}/\Delta m_{B_d}$, 
$A_{CP}^{\text{mix}}(B_d\to J/\psi K_S)$, and $\phi_3$ is very similar as 
the SM. On the other hand, in the non-degenerate case, the 2--3 mixing
in the squark sector is enhanced, and the 1--2 mixing and the 1--3 mixing
are suppressed.
This means that the correlation among $\Delta m_{B_s}/\Delta m_{B_d}$,
$A_{CP}^{\text{mix}}(B_d\to J/\psi K_S)$, and $\phi_3$
may differ from the SM, because of non-negligible SUSY contributions
to the $B_s$--$\bar{B}_s$ mixing.

From this figure, we see that the allowed range of $\phi_3$ depends
on the value of $\Delta m_{B_s}/\Delta m_{B_d}$ in each case. For example,
if $\Delta m_{B_s}/\Delta m_{B_d}$  is consistent with the SM
($\sim 35$), we observe $\phi_3\sim 60^\circ$ in the degenerate case, and
$45^\circ \lesssim \phi_3\lesssim 75^\circ$ in the non-degenerate case.
In the case that $\Delta m_{B_s}/\Delta m_{B_d}$ is larger than the SM,
\emph{e.g.}, $\Delta m_{B_s}/\Delta m_{B_d}\sim 55$, 
$\phi_3$ is $\sim 40^\circ$ in the degenerate case, while 
the allowed range is expected as $45^\circ\lesssim \phi_3\lesssim 60^\circ$
in the non-degenerate case.
This indicates that a $\phi_3$ measurement is important to 
distinguish the two cases.
In Table~\ref{table_utobs_model}, we summarize possible
deviations of $\phi_3$ and $\Delta m_{B_s}$ from the SM in each model.

In the above and the following numerical calculations,
as mentioned in the introduction,
we do not take
the $\tan\beta$-enhanced radiative corrections in the neutral Higgs 
couplings into account. These corrections could be significant for 
the $B-\bar{B}$ mixing amplitudes in the parameter regions of large values of 
$\tan{\beta}$ and small masses of the heavy Higgs bosons, 
because they scale as $(\tan{\beta})^4/(m_{H^{\pm}})^2$ \cite{neut_higgs_bb}. 
Accordingly, it is unlikely that this effect changes our numerical 
results momentously in the region $\tan\beta\lesssim 30$. 
As for $B_d\to \phi K_S$, it is shown by Kane \emph{et al.} \cite{bphiks_susy_rp}
that the contribution from the neutral Higgs exchange diagrams is small.

\begin{figure}
\begin{center}
\includegraphics{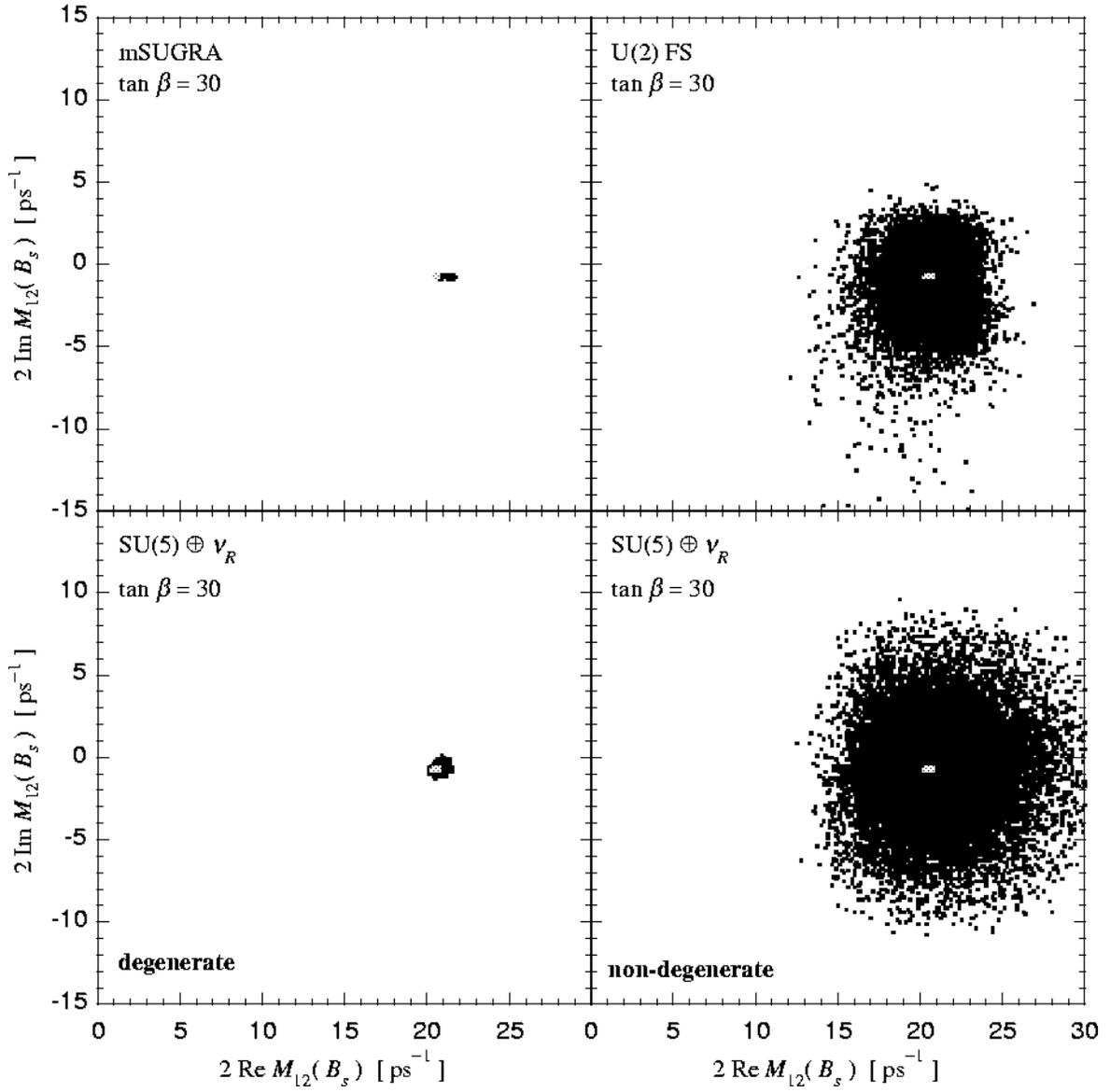}
\end{center}
\caption{
  Real and imaginary parts of $M_{12}(B_s)$.
  The SM value is shown in a light color.
}
\label{fig_m12}
\end{figure}

\begin{table}
\begin{tabular}{c||c|c|c|c}\hline
&mSUGRA&\multicolumn{2}{|c|}{SU(5) SUSY GUT}&U(2)\\\cline{3-4}
&&degenerate&non-degenerate&\\\hline
$\phi_3$&-&$\surd$&-&$\surd$\\
$\Delta m_{B_s}/\Delta m_{B_d}$&-&$\surd$&$\surd$&$\surd$\\ \hline
\end{tabular}
\caption{Possible deviations of $\phi_3$ and 
$\Delta m_{B_s}/\Delta m_{B_d}$ from
values expected in the SM. 
$\surd$ means large deviation.
}
\label{table_utobs_model}
\end{table}

In Fig.~\ref{fig_m12}, we show allowed regions in 
the $\mathrm{Re}M_{12}(B_s)$ and 
$\mathrm{Im}M_{12}(B_s)$ plane for the three models in the case of
$\tan\beta=30$.
In the mSUGRA, the deviation from the SM is 
less than 5\%, and the SUSY contributions to the
complex phase of $M_{12}(B_s)$ are negligible.
In the non-degenerate case of 
the SU(5) SUSY GUT with right-handed neutrinos, the SUSY contribution is
large in
both the real and the imaginary parts of $M_{12}(B_s)$. They can be
as large as 30\% of the SM contribution to $|M_{12}(B_s)|$.
This is in contrast with the degenerate case. 
As seen in Fig.~\ref{fig_m12}, the SUSY contribution to $M_{12}(B_s)$
in the degenerate case is tiny, since the constraint to the parameter 
space imposed by the $\mu\to e\gamma$ branching ratio is very strict.
In the U(2) model, there are SUSY corrections of the order of 20\% 
or larger to $M_{12}(B_s)$.
We have studied the case of $\tan\beta=5$ as well. We have found 
that the allowed
regions of $M_{12}(B_s)$ are similar to those for $\tan\beta=30$. From 
the experimental point of view, $\Delta m_{B_s}=2|M_{12}(B_s)|$ will
be measured by using $B_s$ decays such as $B_s\to D_s\pi$ in hadron $B$
experiments \cite{hadronb}.
The phase of $M_{12}(B_s)$ can be measured
by observing CP violation in $B_s$ decays such as $B_s\to J/\psi \phi$.

\subsubsection{Rare $B$ decays}
Here, we discuss rare $B$ decays in the three models.
We first present SUSY contributions to the Wilson coefficients 
of the dipole operators $C_{7L}$, $C_{7R}$, $C_{8L}$, and $C_{8R}$.
In Table~\ref{table_wc_obs}, we show the relation between
these Wilson coefficients and observables.

\begin{table}
\begin{center}
\begin{tabular}{c|ccc}\hline
&$A_{CP}^{\text{dir}}(B\to X_s\gamma)$&
$A_{CP}^{\text{mix}}(B_d\to M_s\gamma)$&
$A_{CP}^{\text{mix}}(B_d\to \phi K_S)$\\ \hline
$C_{7L}$&$\surd$&$\surd$&-\\
$C_{7R}$&-&$\surd$&-\\
$C_{8L}$&$\surd$&-&$\surd$\\
$C_{8R}$&-&-&$\surd$\\ \hline
\end{tabular}
\end{center}
\caption{The relation between the Wilson coefficients 
and the observables. $\surd$ means that 
the coefficient gives a main contribution to the observables.}
\label{table_wc_obs}
\end{table}

\begin{figure}
\begin{center}
\includegraphics[scale=0.69]{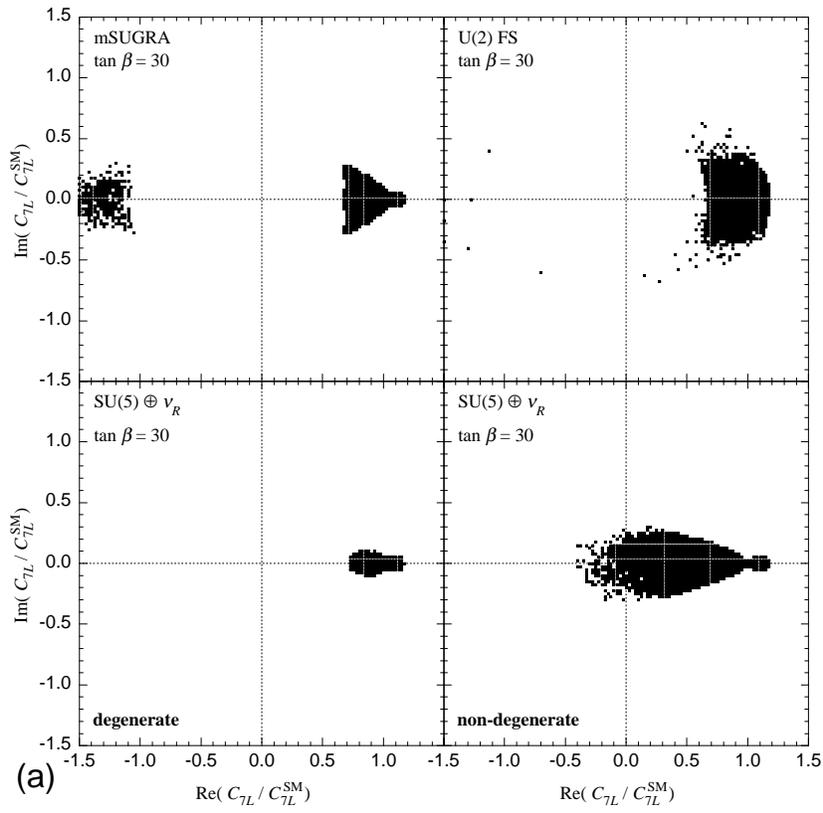}
\includegraphics[scale=0.69]{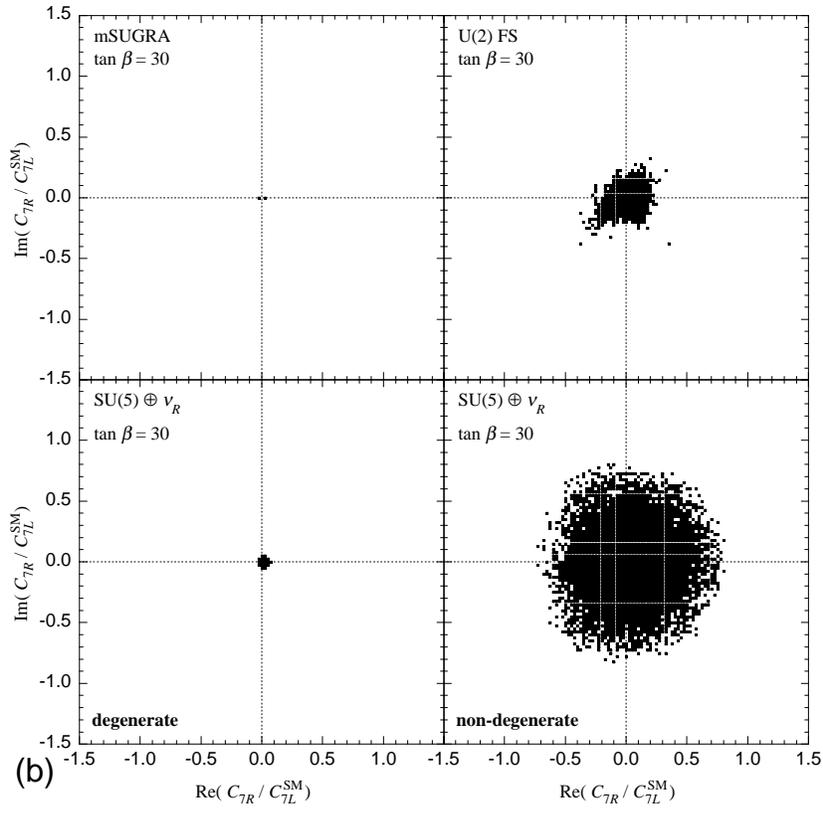}
\end{center}
\caption{
  Wilson coefficients (a) $C_{7L}$ and (b) $C_{7R}$ normalized by the SM
  value of $C_{7L}$.
}
\label{fig_C7}
\end{figure}

The real and imaginary parts of $C_{7L}$ and $C_{7R}$ at the bottom mass scale
divided by the SM value of $C_{7L}$ are plotted in 
Fig.~\ref{fig_C7} for $\tan\beta=30$.
For $\tan\beta=5$, SUSY contributions are
less significant, and we mainly consider 
the $\tan\beta=30$ case in the following.

In the leading order approximation, the branching ratio of $b\to s\gamma$
is proportional to $|C_{7L}|^2+|C_{7R}|^2$. 
In the mSUGRA, however, the SUSY contributions to $C_{7R}$ is very small,
because of no new flavor violation in the right-handed squark sector.
Thus, the $b\to s\gamma$ branching ratio constrains 
$|C_{7L}|$.
In addition, the SUSY contributions to the phase of $C_{7L}$,
which is dominated by the phase $\phi_A$,
is small due to the constraint from the neutron EDM experiment.

In the SU(5) SUSY GUT with right-handed neutrinos, 
the new flavor mixing in the right-handed squark sector is
induced by the MNS matrix and GUT interactions.
SUSY contributions to 
$C_{7L}$ and $C_{7R}$ can be as large as $C_{7L}^{\text{SM}}$.
The EDM constraints
are also strong, and the SUSY contribution to the phase of 
$C_{7L}$ cannot become large as in the mSUGRA.
Since we have introduced no CP phase of the neutrino sector
in this analysis, the SUSY contribution to the phase of $C_{7R}$
mainly comes from the GUT phases $\phi_i^L$. 
Note that $\phi_i^L$'s contribute only to off-diagonal elements of the 
right-handed down squark mass matrix,
and thus, they do not affect the neutron EDM.
The $\mu\to e\gamma$ constraint in the degenerate case is much stronger 
than that in the non-degenerate case as seen in Fig.~\ref{fig_muegamma_su5}.
Therefore, the allowed regions become
much larger in the non-degenerate case.

In the U(2) model, SUSY contributions to 
$C_{7L}$ and $C_{7R}$ can be large because of 
the existence of new flavor mixings in the squark sector.
Though the new contribution to the
phase of $C_{7L}$ is restricted by the neutron EDM constraint 
in this model,
the restriction is weaker than that in the above two models
because the phase of $(m_Q^2)_{23}$
is independent of the phase that contributes to the neutron EDM.
Compared with the non-degenerate case of SU(5) SUSY GUT with 
right-handed neutrinos,
$|(m_D^2)_{23}|$ is suppressed in the U(2) model, and
the SUSY contribution to $|C_{7R}|$ is smaller.

\begin{figure}
\includegraphics[scale=0.69]{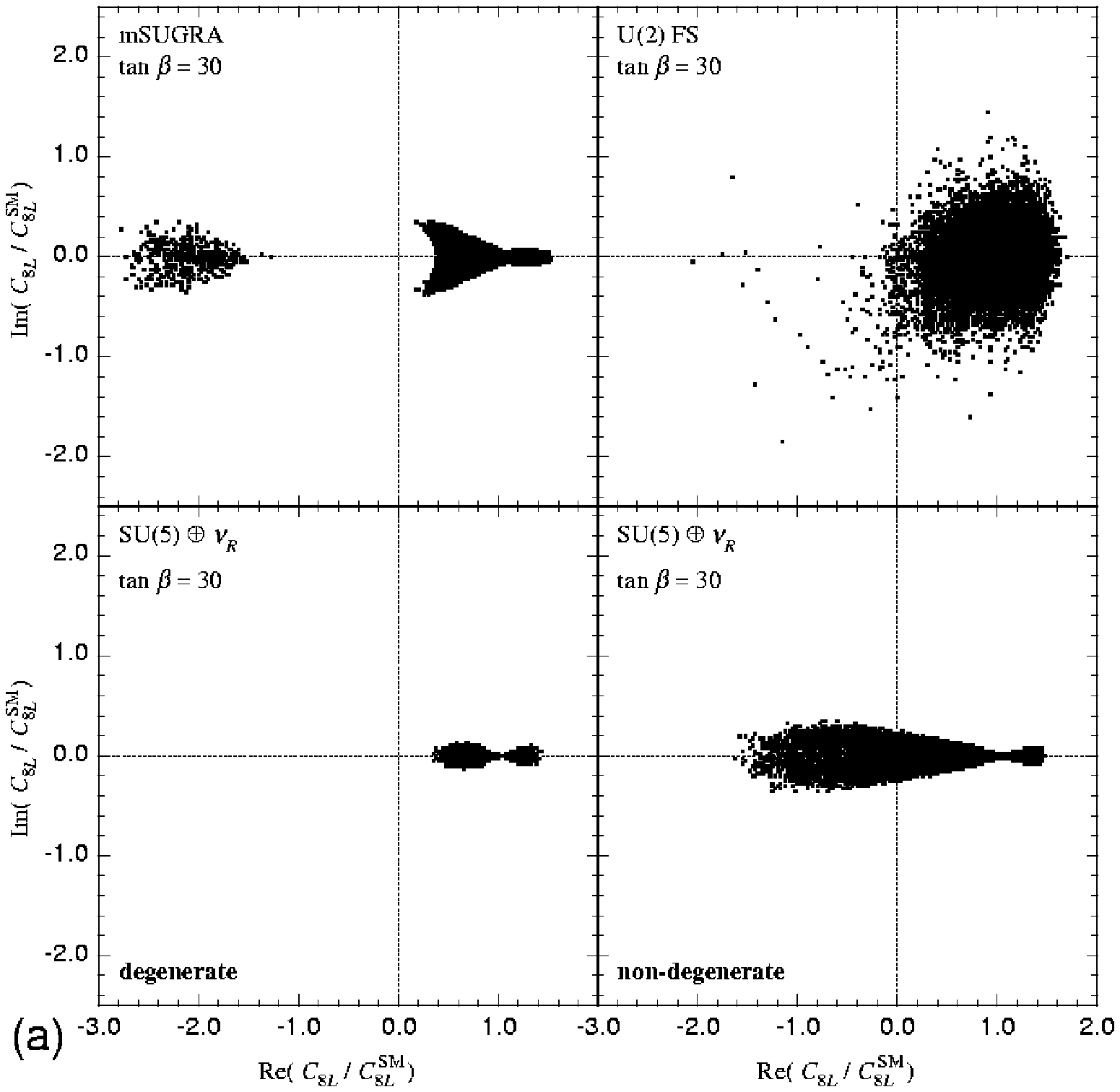}
\includegraphics[scale=0.69]{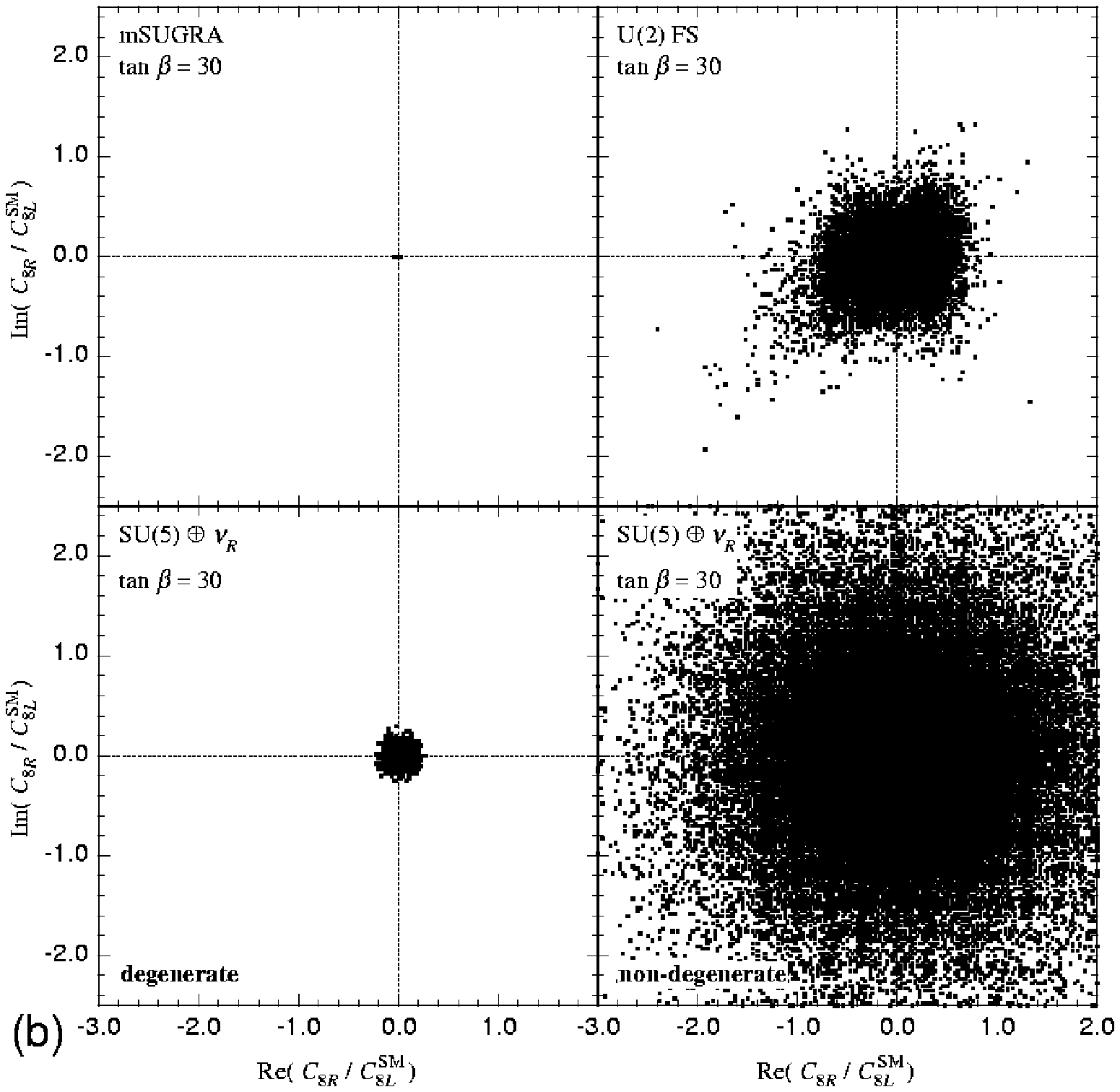}
\caption{
  Wilson coefficients (a) $C_{8L}$ and (b) $C_{8R}$ normalized by the SM
  value of $C_{8L}$.
}
\label{fig_C8}
\end{figure}

The SUSY contributions to $C_{8L}$ and $C_{8R}$ are similar to
those to $C_{7L}$ and $C_{7R}$ in each model as seen in 
Fig.~\ref{fig_C8},
because the flavor mixings and CP phases that determine the SUSY
contributions to $C_{7L}$ and $C_{7R}$ are the same as those contribute
to $C_{8L}$ and $C_{8R}$. This means that the constraints on 
$C_{7L}$ and $C_{7R}$
from $b\to s\gamma$ branching ratio also
restrict predicted values of $C_{8L}$ and $C_{8R}$.
In Table~\ref{table_c78_m12}, we summarize possible SUSY 
contributions to Wilson coefficients $C_7$'s and $C_8$'s 
and $M_{12}(B_s)$ in each models.

\begin{table}
\begin{tabular}{c||c|c|c|c}\hline
&mSUGRA&\multicolumn{2}{|c|}{SU(5) SUSY GUT}&U(2)\\\cline{3-4}
&&degenerate&non-degenerate&\\\hline
$|C_{7,8L}|$&-&-&$\surd\surd$&$\surd\surd$\\
$|C_{7,8R}|$&-&$\surd$&$\surd\surd$&$\surd\surd$\\
$\mathrm{arg}C_{7,8L}$&$\surd$&-&$\surd$&$\surd\surd$\\
$\mathrm{arg}C_{7,8R}$&-&$\surd$&$\surd\surd$&$\surd\surd$\\
$M_{12}(B_s)$&-&-&$\surd\surd$&$\surd\surd$\\ \hline
\end{tabular}
\caption{Possible SUSY contributions to Wilson coefficients
$C_{7}$'s and $C_{8}$'s and
$M_{12}(B_s)$ in each model. 
$\surd$ means non-negligible deviation from the SM,
and $\surd\surd$ denotes large SUSY contributions.}
\label{table_c78_m12}
\end{table}

\begin{figure}
\begin{center}
\includegraphics[scale=0.68]{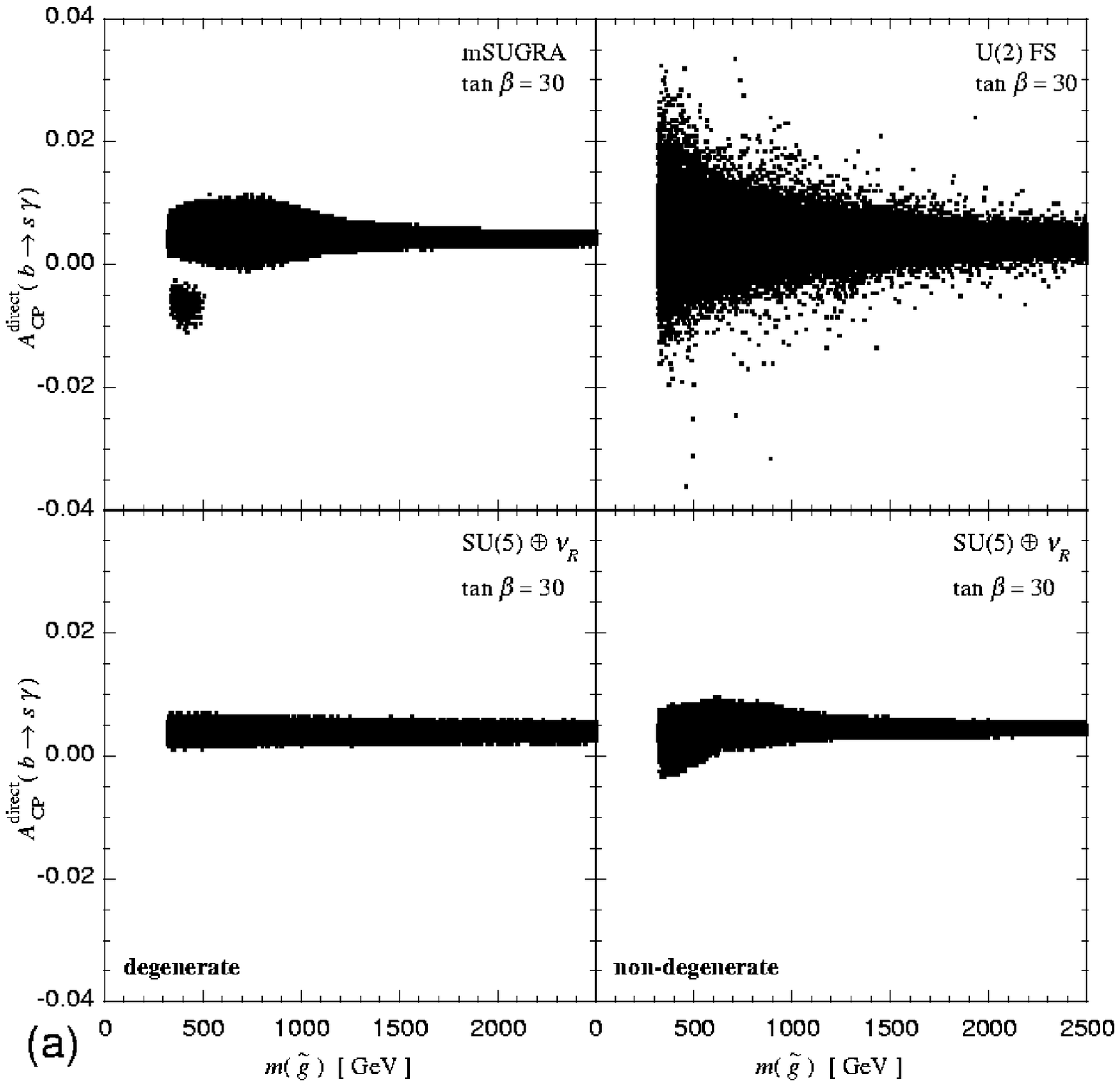}
\includegraphics[scale=0.68]{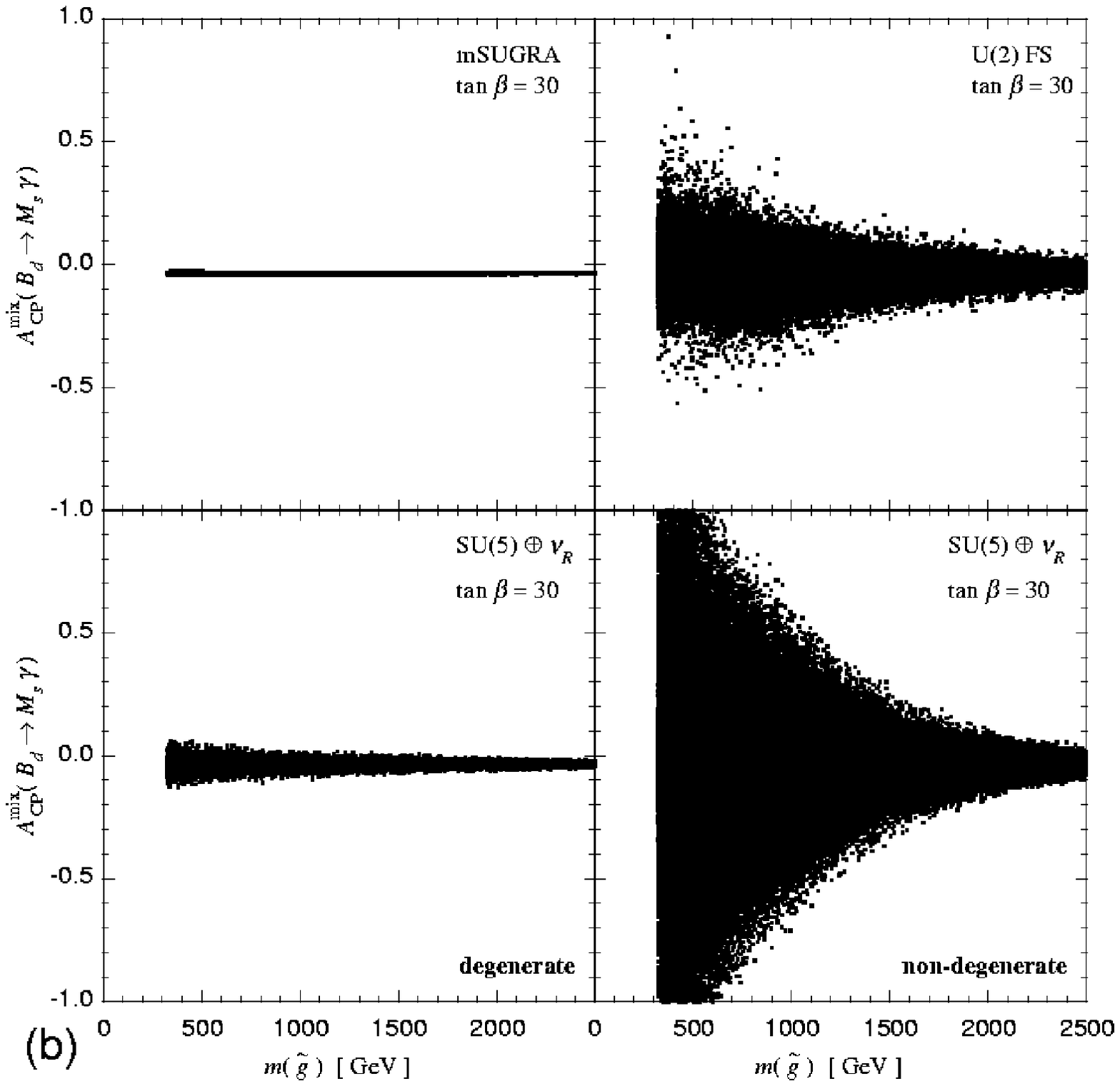}
\end{center}
\caption{
  (a) The direct CP asymmetry in $b\to s\gamma$, 
  and (b) the mixing-induced CP asymmetry in $B_d\to M_s\gamma$
  as functions of the gluino mass.
}
\label{fig_phaseinbsgamma}
\end{figure}

In Fig.~\ref{fig_phaseinbsgamma}, we plot
$A_{CP}^{\text{dir}}(B\to X_s\gamma)$ and 
$A_{CP}^{\text{mix}}(B\to M_s\gamma)$ versus the gluino mass
for $\tan\beta=30$.
$A_{CP}^{\text{dir}}(B\to X_s\gamma)$
essentially comes from the imaginary part of the
interference terms between 
$C_{7L}$ and $C_{2L}$, and
$C_{8L}$ and $C_{2L}$, because $C_{2R}$ is negligible.
Therefore, SUSY contributions to $A_{CP}^{\text{dir}}(B\to X_s\gamma)$ 
is constrained by the neutron EDM, and $|A_{CP}^{\text{dir}}(B\to X_s\gamma)|$
is at most $\sim 1\%$ in the mSUGRA
and the SU(5) SUSY GUT with right-handed neutrinos. 
$|A_{CP}^{\text{dir}}(B\to X_s\gamma)|$ can be
as large as 3\% in the U(2) model.
The SM prediction is about 0.5\% \cite{directcpbsgamma}.
On the other hand, the mixing induced CP asymmetry in $B_d\to M_s\gamma$ 
depends on $C_{7L}$ and $C_{7R}$.
Although the SUSY contribution to $C_{7L}$ cannot be negligible 
in the models that we are studying, the deviation from the SM essentially 
comes from $C_{7R}$. Therefore, the  SUSY effect can
become larger in the SU(5) SUSY GUT with right-handed neutrinos
and in the U(2) model
compared with the mSUGRA model.
In the mSUGRA model, $|A_{CP}^{\text{mix}}(B\to M_s\gamma)|$ is
at a level of 1\%, which is similar to the value of the 
SM \cite{mixcpbsgamma}.
In the non-degenerate case of the SU(5) SUSY GUT with right-handed neutrinos, 
$|A_{CP}^{\text{mix}}(B_d\to M_s\gamma)|$ can be maximal, while in the
degenerate case, $|A_{CP}^{\text{mix}}(B_d\to M_s\gamma)|$
can be as large as 0.1.
In the U(2) model, we find that 
$|A_{CP}^{\text{mix}}(B\to M_s\gamma)|$ could be as large as
0.5.

\begin{figure}
\begin{center}
\includegraphics{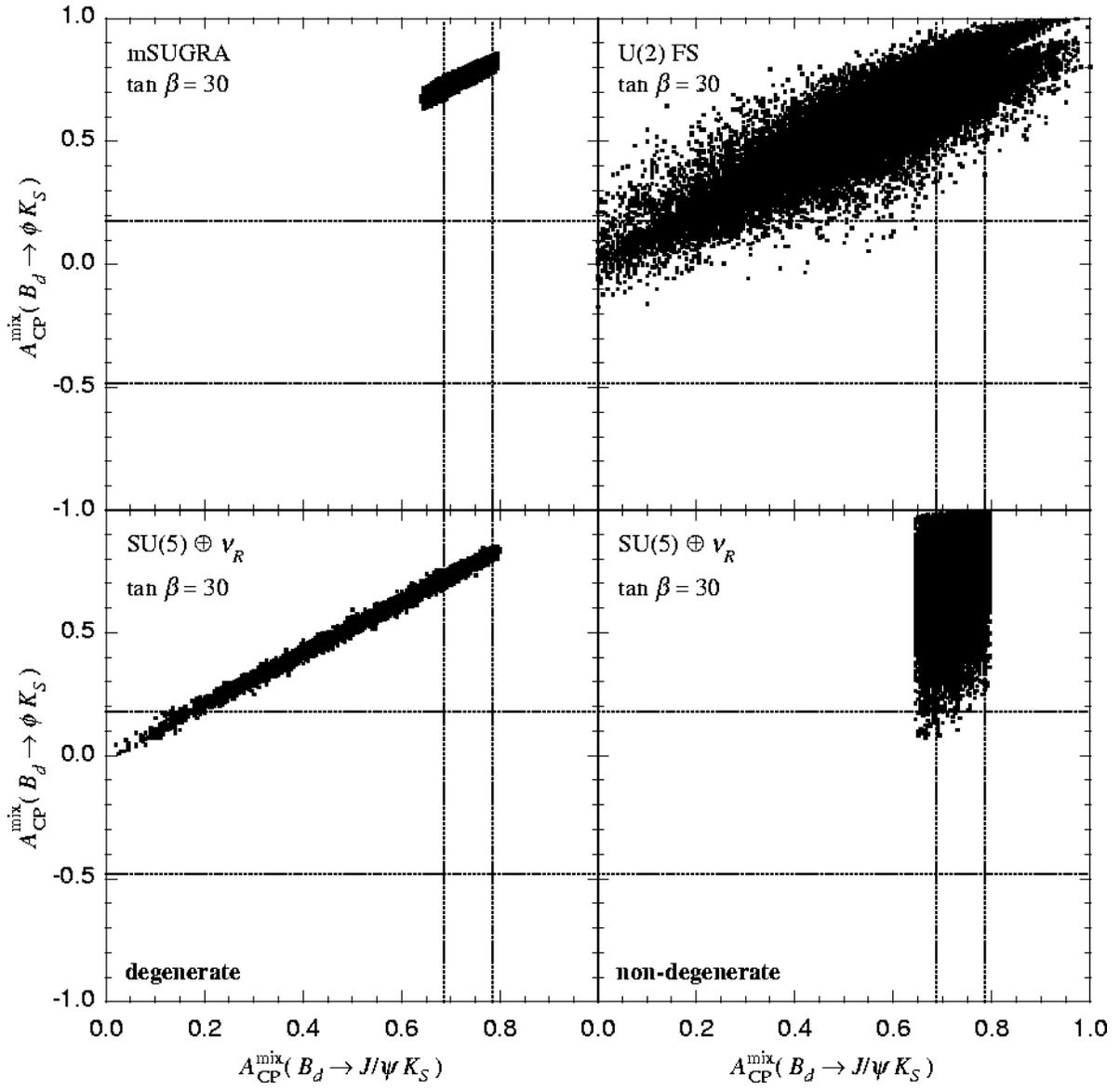}
\end{center}
\caption{
  The correlation between the mixing-induced CP asymmetries in
  $B_d\to \phi K_S$ and $B_d\to J/\psi K_S$.
  The vertical and horizontal dotted lines show the $1\sigma$ ranges of
  experimental values.
  In this plot, the experimental constraint of
  $A_{CP}^{\text{mix}}(B_d\to J/\psi K_S)$ is not imposed.
}
\label{fig_btophik}
\end{figure}

In Fig.~\ref{fig_btophik}, we show the correlation between 
$A_{CP}^{\text{mix}}(B_d\to \phi K_S)$ and 
$A_{CP}^{\text{mix}}(B_d\to J/\psi K_S)$ for $\tan\beta=30$.
In the SM, 
$A_{CP}^{\text{mix}}(B_d\to \phi K_S)=A_{CP}^{\text{mix}}(B_d\to J/\psi K_S)$
is satisfied.
As mentioned in Sec.~\ref{sec_processes}, the SM contribution is dominant
in $C_{LL}$, $C_{LR}^{(1)}$, and $C_{LR}^{(2)}$
due to the QCD correction between the electroweak scale and 
the bottom mass scale.
Thus, SUSY contributes to 
$A_{CP}^{\text{mix}}(B_d\to \phi K_S)$ mainly through $C_8$'s.

In the mSUGRA, we see that the SM relation 
$A_{CP}^{\text{mix}}(B_d\to \phi K_S)=A_{CP}^{\text{mix}}(B_d\to J/\psi K_S)$
approximately holds, and 
the deviation from the SM in
$A_{CP}^{\text{mix}}(B\to J/\psi K_S)$ is 
less than 10\% as seen in Ref.~\cite{gosst1}. Therefore,
$A_{CP}^{\text{mix}}(B\to \phi K_S)$ is 
almost the same as that in the SM.

In the non-degenerate case of the SU(5) SUSY GUT with right-handed neutrinos,
$A_{CP}^{\text{mix}}(B_d\to \phi K_S)$
can substantially differ from the value in the SM and 
may be smaller than 0.1,
because of the large SUSY contributions to $C_{8R}$.
On the other hand, in the degenerate case, the SM
relation 
$A_{CP}^{\text{mix}}(B_d\to \phi K_S)=A_{CP}^{\text{mix}}(B_d\to J/\psi K_S)$
is satisfied. Accordingly, the value of $A_{CP}^{\text{mix}}(B_d\to \phi K_S)$
is restricted by the experimental result on
$A_{CP}^{\text{mix}}(B_d\to J/\psi K_S)$.

In the U(2) model, 
$A_{CP}^{\text{mix}}(B\to \phi K_S)$ can deviate from 
the SM prediction because of SUSY contributions to $C_{8L}$ and $C_{8R}$.
The experimental result of $A_{CP}^{\text{mix}}(B_d\to J/\psi K_S)$
implies that $A_{CP}^{\text{mix}}(B_d\to \phi K_S)$ lies between 
0.3 and 1.0.

\begin{figure}
\includegraphics{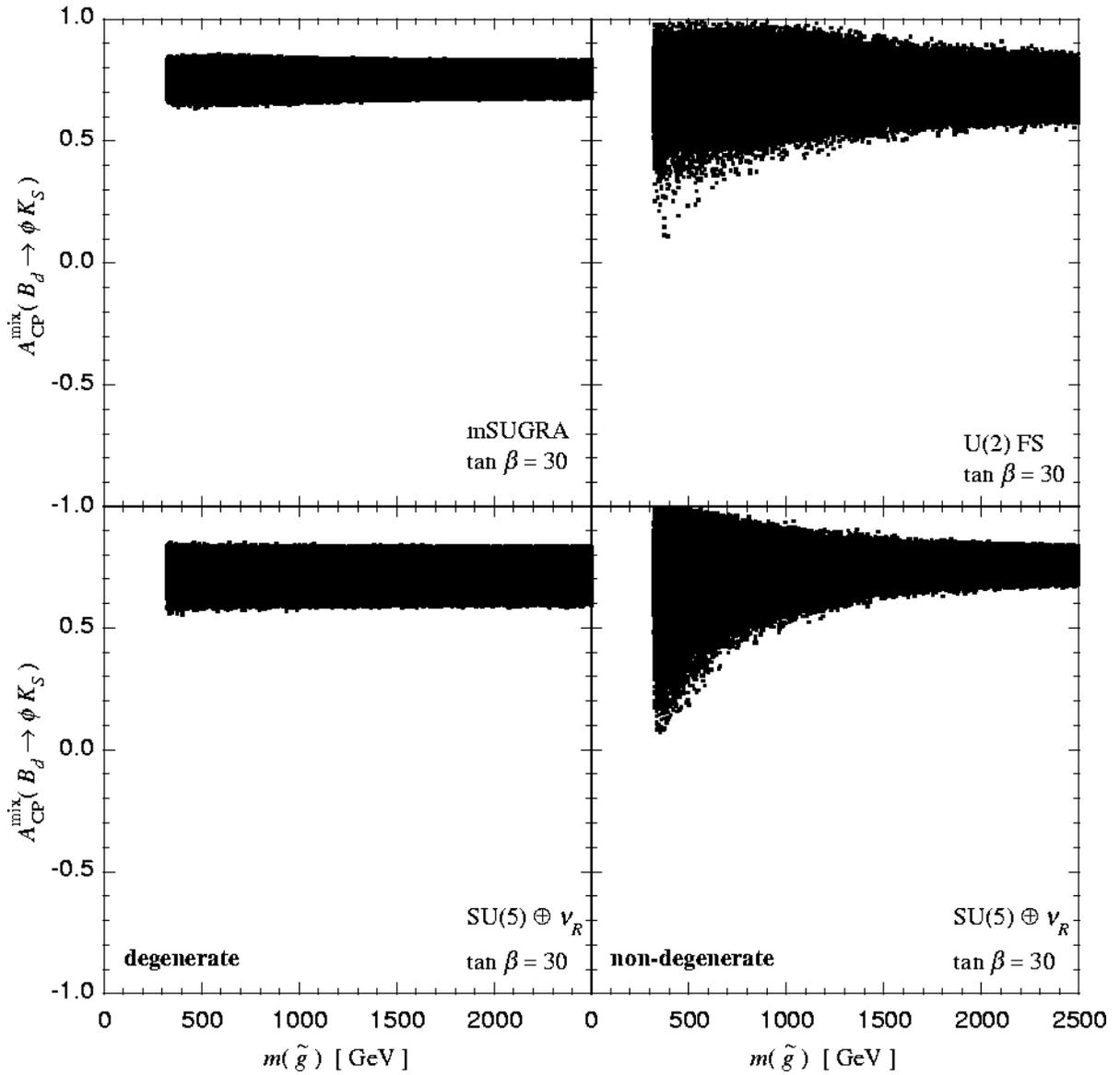}
\caption{
  The mixing-induced CP asymmetry in $B_d\to \phi K_S$ as a function of
  the gluino mass.
}
\label{fig_btophik-gno}
\end{figure}

In Fig.~\ref{fig_btophik-gno}, we show $A_{CP}^{\text{mix}}(B_d\to \phi K_S)$
as a function of the gluino mass.
In the mSUGRA and the degenerate case of the SU(5) SUSY GUT with right-handed
neutrinos, the SUSY effect is almost negligible, and we see virtually no 
dependence on the gluino mass. 
On the other hand, in the degenerate case of the SU(5) SUSY GUT with
right-handed neutrinos and the U(2) model, the SUSY contribution
is significant, in particular for the smaller mass of the gluino.

\begin{figure}
\includegraphics{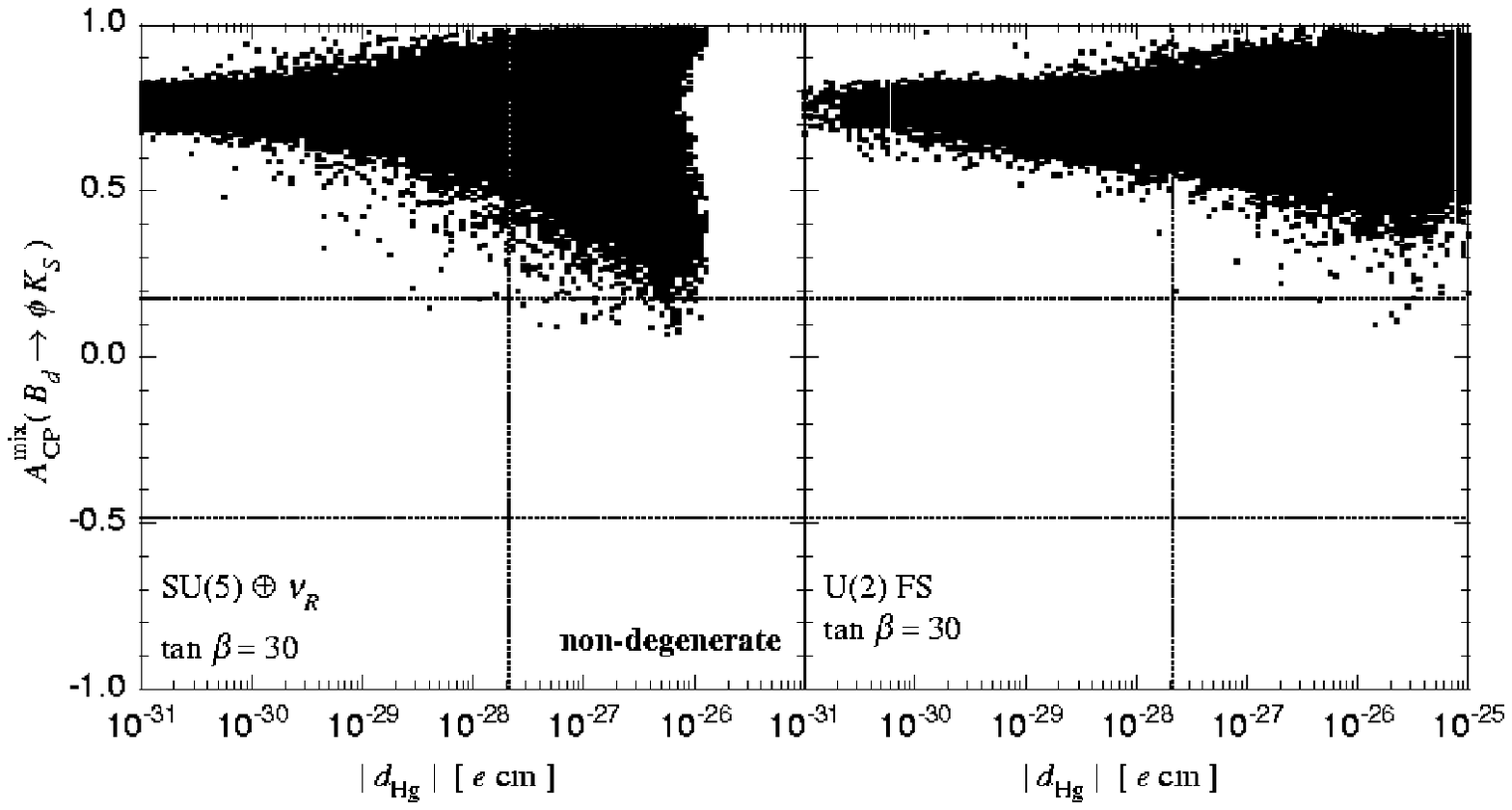}
\caption{
The correlation between the mixing-induced CP asymmetry in $B_d\to \phi K_S$ 
and the $^{199}$Hg EDM in the non-degenerate case of the SU(5) SUSY GUT with 
right-handed neutrinos and the U(2) model.
}
\label{fig_dHg-phiKS}
\end{figure}
\begin{figure}
\includegraphics{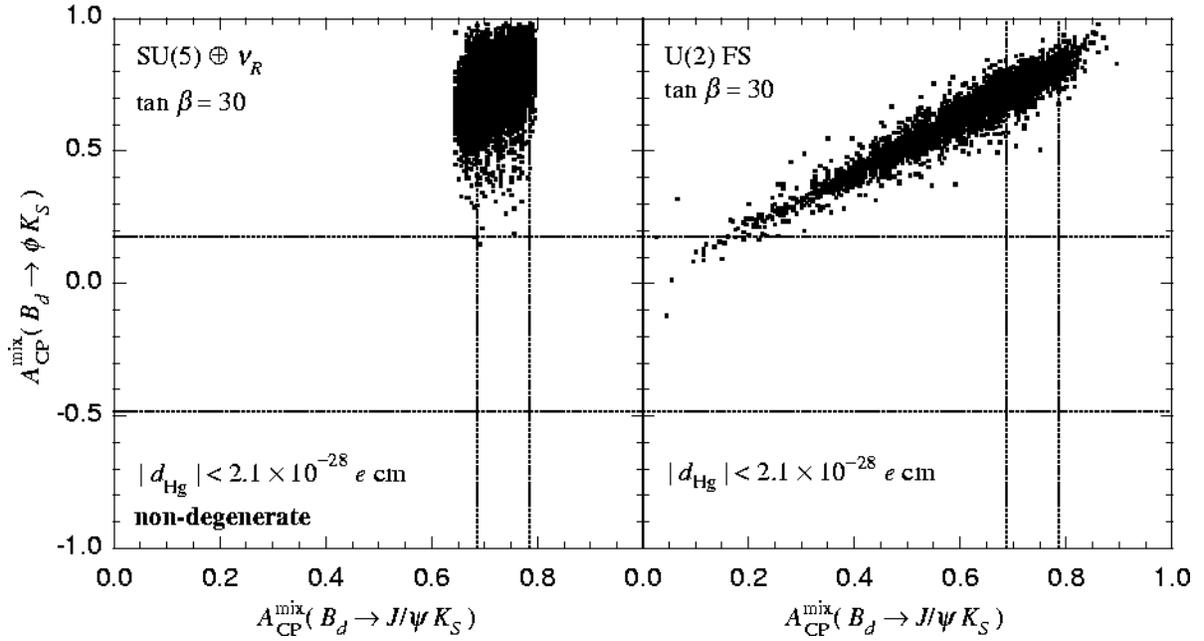}
\caption{
  The correlation between the mixing-induced CP asymmetries in
  $B_d\to \phi K_S$ and $B_d\to J/\psi K_S$ under the constraint
  of the $^{199}$Hg EDM in the non-degenerate case of the SU(5) SUGY GUT
  with right-handed neutrinos and the U(2) model.
}
\label{fig_psiKS-phiKS_Hg}
\end{figure}

Recently, it has been pointed out that the contribution of the chromo-EDM 
of the strange quark to the EDM of $^{199}$Hg is devastating in 
SUSY models with a large 2--3 mixing in the right-handed squark 
sector, and that a large deviation of $A_{CP}^{\text{mix}}(B_d\to \phi K_S)$ 
from the SM prediction is unlikely provided that the experimental upper 
bound \cite{exp_edm_hg} on the EDM of $^{199}$Hg is imposed \cite{edm_hg_phiK}. 
Although the theoretical estimate \cite{edm_hg} of the $^{199}$Hg EDM due to 
the chromo-EDM of the light quarks suffers from relatively large
uncertainties, it is probable that the $^{199}$Hg EDM constraint severely
restricts flavor mixings and CP violation in some SUSY models, 
and thus constricts their flavor and CP signals such as 
$A_{CP}^{\text{mix}}(B_d\to \phi K_S)$ in general. 

Among the models considered in the present work, effects of 
the $^{199}$Hg EDM constraint are non-negligible in the non-degenerate 
case of the SU(5) SUSY GUT with right-handed neutrinos and the U(2) model. 
We show the correlation between the $^{199}$Hg EDM and 
$A_{CP}^{\text{mix}}(B_d\to \phi K_S)$ in these models in 
Fig.~\ref{fig_dHg-phiKS}. 
In our numerical calculation of the $^{199}$Hg EDM,
contributions of all the three light flavors are included.
Fig.~\ref{fig_dHg-phiKS} illustrates how a flavor
signal is tightened if the $^{199}$Hg EDM constraint is applied.
In Fig.~\ref{fig_psiKS-phiKS_Hg}, the correlation between 
$A_{CP}^{\text{mix}}(B_d\to J/\psi K_S)$ and 
$A_{CP}^{\text{mix}}(B_d\to \phi K_S)$ under the constraint of 
the $^{199}$Hg EDM is shown. Comparing this figure with Fig.\ref{fig_btophik},
we see that the $^{199}$Hg EDM constraint
certainly restricts the possible deviation from the SM prediction.
At the same time, however, our detailed calculation shows that
there still remains some parameter region in which 
$A_{CP}^{\text{mix}}(B_d\to \phi K_S)$ is significantly different from
$A_{CP}^{\text{mix}}(B_d\to J/\psi K_S)$ in these models.
A similar argument is applied to both $A_{CP}^{\text{dir}}(B\to X_s\gamma)$ 
and $A_{CP}^{\text{mix}}(B\to M_s\gamma)$.

In Table~\ref{table_obs_model}, 
we summarize the significance of SUSY contributions on the 
CP asymmetries that we have considered.
In this table, we see the possibility to distinguish the three models
in $B$ experiments.

\begin{table}
\begin{tabular}{c||c|c|c|c}\hline
&mSUGRA&\multicolumn{2}{|c|}{SU(5) SUSY GUT}&U(2)\\\cline{3-4}
&&degenerate&non-degenerate&\\\hline
$A^{\text{dir}}_{CP}(B\to X_s\gamma)$&$\surd$&-&$\surd$&$\surd\surd$\\
$A^{\text{mix}}_{CP}(B\to M_s\gamma)$&-&$\surd$&$\surd\surd$&$\surd\surd$\\
$A_{CP}^{\text{mix}}(B\to \phi K_S)$&-&-&$\surd\surd$&$\surd\surd$\\ \hline
\end{tabular}
\caption{Significance of SUSY contributions to the CP asymmetries
in each model.
$\surd$ means non-negligible deviation from the SM, and
$\surd\surd$ means large SUSY contributions.}
\label{table_obs_model}
\end{table}

As we stressed in the introduction, the purpose of this work is to 
demonstrate that identifying patterns of deviations from the SM
predictions is useful to distinguish different origins of the 
SUSY breaking sector. From this point of view, combined with 
the analysis of the unitarity triangle\cite{gosst1}, 
we can make the following observations:
\begin{itemize}
\item Deviations from the SM predictions in the unitarity
triangle and rare decays are small in the mSUGRA model, except for
some sizable contributions in the direct CP violation in the $b\to s\gamma$
process. Note that this conclusion may not hold in a 
particularly large value of $\tan{\beta}\sim 60$ due to the Higgs exchange 
effects \cite{neut_higgs_bb}.
\item The pattern of the deviations from the SM
depends on the right-handed neutrino mass matrix in
the SU(5) SUSY GUT with right-handed neutrinos.
In the degenerate case, flavor mixing signals between the 1--2 generations
become large. This appears as inconsistency between the measured value
of $\epsilon_K$ and the $B$ meson unitarity triangle, although the unitarity
triangle is closed among $B$ meson observables. The rare decay processes
induced by the $b$--$s$ transition do not show large deviations, but
the branching ratio of $\mu \to e \gamma$ process can be just below the 
present experimental bound. This is expected to be a generic feature of
SU(5) SUSY GUT with right-handed neutrinos.    
\item In a specific parameter choice of the ``non-degenerate'' case, in which
the $\mu \to e \gamma$ constraint is relaxed, the flavor signals between
2--3 generations are expected to be sizable. This includes the 
mixing-induced CP asymmetry in $B_d\rightarrow M_s\gamma$ and
$B_d\rightarrow\phi K_S$. The direct CP asymmetry in the $b\rightarrow s\gamma$
process, on the other hand, does not show a large deviation.  
\item Various new physics signals in the consistency test of the unitarity 
triangle and rare decay process are expected in the MSSM with U(2)
flavor symmetry.
\end{itemize}
In this way, we can expect different sizes and patterns of new physics 
signals in the above models. These are crucial in pointing toward a specific
model from flavor physics.

\section{Conclusions}

In order to seek the possibility to distinguish different
SUSY models with $B$ physics experiments, we have studied
rare $B$ decays related to the $b\rightarrow s$ transition
combining with the unitarity triangle analysis in three SUSY models.
These models, namely the mSUGRA,
the SU(5) SUSY GUT with right-handed neutrinos, and the
U(2) flavor symmetry model, are different in character
with respect to flavor structures of their SUSY breaking sectors.
We have considered two different cases in regard to the mass
spectrum of the right-handed neutrinos in the SU(5) SUSY
GUT with right-handed neutrinos.

In the unitarity triangle analysis, we have studied consequences of SUSY to 
$A_{CP}^{\text{mix}}(B_d\to J/\psi K_S)$,
$\Delta m_{B_s}/\Delta m_{B_d}$, and $\phi_3$.
Our results are summarized in Table~\ref{table_utobs_model} 
and Fig.~\ref{fig_dmbsd_acppsiks}.
It could be possible to distinguish the three models
by precisely measuring $\Delta m_{B_s}$ and $\phi_3$
in future $B$ experiments.

As for rare $B$ decays, we have explored SUSY effects to the direct 
CP asymmetry in
$b\rightarrow s\gamma$, the mixing induced CP asymmetry in
$B_d\rightarrow M_s\gamma$, and the CP asymmetry in
$B_d\rightarrow\phi K_S$ 
in the three models. The results are summarized in Tables
\ref{table_c78_m12} and \ref{table_obs_model}. 
Table~\ref{table_c78_m12} shows the relative importance of SUSY contributions
to the theoretically interesting Wilson coefficients related to
the $b\rightarrow s$ transitions and the $B_s$--$\bar B_s$
mixing amplitude. The significance of SUSY effects to
the CP asymmetries is indicated in Table~\ref{table_obs_model}.

The new flavor signals in the mSUGRA and the degenerate case of
the SU(5) SUSY GUT are relatively limited in the $b\rightarrow s$
rare decays considered in the present work. To detect these signals,
typically a few percent, we may need an ultimate $B$ experiment.

On the other hand, the non-degenerate case of the SU(5) SUSY GUT
exhibits quite attractive flavor signals in $B_d\rightarrow M_s\gamma$
and $B_d\rightarrow\phi K_S$ as seen in Table~\ref{table_obs_model}. 
We have also observed that
the U(2) model predicts significant deviations from the SM in the
$b\rightarrow s$ rare decays as well as the unitarity triangle analysis.
So far, both Belle and BaBar experiments have collected copious $B$ decays,
and they are expected to go well continuously.
Thus, more $B_d\to \phi K_S$ and related events will be obtained in
near future. Moreover, both KEK and SLAC plan to upgrade their $B$ 
factories.
Therefore, the above flavor signals may well be in
the reach of the present and foreseeable future $B$ experiments.

Combining the above observation with the results in our previous work,
we conclude that the study of the unitarity triangle and rare $B$ decays
could discriminate several SUSY models that have different flavor
structures in their SUSY breaking sectors. 
Such a study will play important roles, even if SUSY particles
are found at future experiments at the energy frontier such as LHC.
Although the spectrum of SUSY particles will be determined at
LHC and a future $e^+e^-$ linear collider, most of information concerning
the flavor mixing of the squark sector is expected to
come from the super $B$ factory and hadron $B$ experiments.
Since the flavor structure
of the SUSY breaking provides us with an important clue 
to the origin of the SUSY breaking mechanism and interactions at very
high energy scales, $B$ physics will be essential for clarifying a
whole picture of the SUSY model.

\acknowledgments{
The work of Y.O. was supported
in part by a Grant-in-Aid of the Ministry of Education, Culture, Sports,
Science, and Technology, Government of Japan (No.~13640309).
The work of Y.S. was supported in part by a Grant-in-Aid of
the Ministry of Education, Culture, Sports, Science and
Technology, Government of Japan (No.~13001292).
The work of M.T. was supported in part by a Grant-in-Aid of
the Ministry of Education, Culture, Sports, Science and
Technology, Government of Japan (No.~14046212 and No.~15540272).
The work of T.S. was supported in part by Research Fellowship of the
Japan Society for the Promotion of Science (JSPS) for
Young Scientists (No.15-03927).
}
 
\end{document}